\documentclass[pra,twocolumn,aps,showpacs,floatfix,superscriptaddress]{revtex4-1}
\usepackage{graphicx}
\usepackage{rotating}
\usepackage{amsmath}
\usepackage{bbm}
\usepackage{subfigure}
\usepackage{booktabs}
\usepackage{multirow}
\usepackage{color}

\renewcommand{\v}[1]{{\bf #1}}

\newcommand{\be}{\begin{equation}}
\newcommand{\ee}{\end{equation}}
\newcommand{\bea}{\begin{eqnarray}}
\newcommand{\eea}{\end{eqnarray}}
\newcommand{{\br}}{\bf r}

\graphicspath{
  {./}
  {figures/pdf/}
  {figures/eps/}
  {figures/jpg/}
}

\begin{document}
\title{Conditions for Describing Triplet States in Reduced Density Matrix Functional Theory}
\author{Iris Theophilou}
\affiliation{Max Planck Institute for the Structure and Dynamics of Matter, 
Luruper Chaussee 149, 22761 Hamburg, Germany}
\affiliation{Peter-Gr\"unberg Institut and Institute for Advanced Simulation,
Forschungszentrum J\"ulich, D-52425 J\"ulich, Germany}
\email{iris.theophilou@mpsd.mpg.de}
\author{Nektarios N.\ Lathiotakis}
\affiliation{Theoretical and Physical Chemistry Institute, National Hellenic 
Research Foundation, Vass.\  Constantinou 48, GR-11635 Athens, Greece}
\affiliation{Max-Planck-Institut f\"ur Mikrostrukturphysik, Weinberg 2, D-06120 Halle (Saale), Germany}
\author{Nicole Helbig}
\affiliation{Peter-Gr\"unberg Institut and Institute for Advanced Simulation,
Forschungszentrum J\"ulich, D-52425 J\"ulich, Germany}

\begin{abstract}  
\noindent

We consider necessary conditions for the one-body-reduced density matrix (1RDM) to correspond to a 
triplet wave-function of a 2-electron system. The conditions concern the occupation numbers 
and are different for the high spin projections, $S_z=\pm 1$, and the $S_z=0$ projection. Hence, they can be used to test if an approximate 1RDM functional yields the same energies for both projections. We employ these conditions in reduced density matrix functional theory calculations for the triplet excitations of two electron systems. In addition, we propose that these conditions can be used in the calculation of triplet states of systems with more than two electrons by restricting the active space. We assess this procedure in calculations for a few atomic and molecular systems. We show that the quality of the optimal 1RDMs improves by applying the conditions in all the cases we studied.
\end{abstract}

\date{\today}

\maketitle


\section{Introduction\label{sec:intro}}

The main focus of reduced density matrix functional theory (RDMFT) \cite{G1975}, a framework where the one-body reduced density matrix (1RDM) plays the role of the fundamental variable, has been the proper description of ground-state singlet states. Little has been done on developing a RDMFT treatment of doublet or triplet states \cite{LHG2005, Piris1, RP2011, Piris_Leiva_spin}. The extension of the theory for such states can follow two different directions which can also be combined. The first concerns the development of approximate functionals or the extension of existing ones to describe such states. The second is the derivation of additional conditions to restrict the minimization of the existing functionals to the domain of 1RDMs that correspond to a prescribed spin state. The present work is a step in the second direction.

Since the many-electron problem, in general, cannot be solved exactly, several approximations were introduced where the total electronic energy is expressed as a functional of a density or density matrix. In this way, one switches from calculating the many-body state to calculating quantities like the density in density functional theory (DFT) \cite{HK1964} or the 1RDM in RDMFT.
RDMFT \cite{G1975,Pernal2016} got significant attention in the last 20 years as an alternative to DFT. Several approximations have been introduced \cite{M1984,GU1998,BB2002,GPB2005,LHZG2009,P2014,KP2014,RPGB2008,ML2008,SDLG2008,SSSLG2015,LSDEMG2009,P2006,PMLU2010,PLRMU2011,S1999,R2002,RX2007,LHRG2014}
with promising results in cases like molecular dissociation \cite{BB2002,GPB2005,LHZG2009,P2014,KP2014,MPRLU2011} or the gaps of periodic systems \cite{HLAG2007,SDLG2008,Letc2010,SSSLG2015}, where the results of basic DFT functionals are not satisfactory. Among these approximations, a central position is held by the M\"uller functional \cite{M1984,BB2002} which was found to overcorrelate substantially. Its inaccurate reconstruction of the 2RDM in terms of the 1RDM is manifested by the violation of the positivity of the 2-particle density, which was already recognized by M\"uller himself. However, the M\"uller functional served as a starting point for further improvements. Several other functionals were introduced \cite{Pernal2016} aiming to correct its overcorrelation: the functional of Goedecker and Umrigar \cite{GU1998} the BBC$n$ ($n$ = 1,2,3) \cite{GPB2005,RPGB2008,LHZG2009}, the approximation of Marques and Lathiotakis \cite{ML2008} and the Power functional \cite{SDLG2008,LSDEMG2009,Constraints_Power,SSSLG2015}. In a different fashion, PNOF$n$ approximations, $n$ = 1,$\cdots$6 \cite{P2006,PMLU2010,PLRMU2011,P2014}, and the theory of the antisymmetrized product of strongly orthogonal geminals (APSG) \cite{S1999,R2002,RX2007,KP2014} were developed focusing on improving the reconstruction of the 2RDM. 

In any RDMFT calculation, one searches for the 1RDM that minimizes the total energy functional. However, the search has to be restricted in the  domain of functions (trial 1RDMs) that satisfy certain constraints, known as $N$-representability conditions, which guarantee that the optimal 1RDM corresponds to a fermionic system. Given the exact ground-state total energy functional, the ensemble $N$-representability conditions of Coleman \cite{Coleman_1963} are sufficient to ensure that one finds the 1RDM that corresponds to the nondegenerate ground-state wave function. The reason is that any ensemble of pure states would always include excited states and would lead to a higher total energy. For approximate functionals, on the other hand, the ensemble conditions do not guarantee that the minimizing 1RDM could be obtained from a many-body fermionic wave function. The class of spin-compensated systems with time-reversal symmetry is a notable exception, since, in that case, the conditions for pure-state $N$-representability collapse to the ensemble conditions \cite{S1966}. The necessary and sufficient conditions for pure-state $N$-representability, also called generalized Pauli constraints, have only recently been discussed and explicitly expressed for systems with a small number of particles and specific finite sizes of the Hilbert space \cite{ThesisAltunbulak, Klyachko_Math, Borland-Dennis, PRL_pure_cond,Carlos1, Carlos2, Schilling_2015}. 
Recently, it has been demonstrated that with enforcing only the ensemble conditions in a RDMFT calculation for open-shell systems, the pure-state conditions will be violated for many functionals of the 1RDM \cite{TLMH2015}. Hence, the enforcement of the pure-state conditions leads to a different solution. We should mention that the number of pure-state conditions explodes as the number of electrons and the dimension of the Hilbert space increase and their consideration in a minimization procedure becomes a very difficult task.

In many cases, the many-body Hamiltonian commutes with the total spin and the spin projection in any particular direction. As a result, one can choose the solutions of the many-body Schr\"odinger equation to be eigenstates of the Hamiltonian, $\hat{\mathbf{S}}^2$, and $\hat{S}_z$ simultaneously. Typical cases where the Hamiltonian does not commute with the spin operators include the application of nonuniform magnetic fields to the system or the inclusion of spin-orbit coupling. However, we are not considering such cases in this work. Thus, for the cases discussed here, it is desirable that the optimal 1RDMs correspond to eigenstates of the spin operators as well. While the pure-state conditions ensure that there exists a many-body wave function corresponding to a given 1RDM, there is generally no guarantee that this many-body state is an eigenstate of any spin operator or even corresponds to a specific expectation value of it. Hence, the question arises if one can find constraints such that the solutions preserve the symmetries of the original many-body Hamiltonian, i.e., whether the optimal 1RDM which one finds in a RDMFT calculation corresponds to a many-body state with a prescribed, specific total spin and $S_z$. For the $z$-component of the spin, one typically constrains separately the number of up and down electrons in the system \cite{LHG2005} to the correct integer values. 
The expectation value of $\hat{S}_z$ is then given by $(N_\uparrow-N_\downarrow)/2$ (atomic units are used throughout this paper unless explicitly stated otherwise). However, this constraint guarantees only that the expectation value of $\hat{S}_z$ has the correct, prescribed value. Therefore, it is a necessary condition for the 1RDM to correspond to an eigenstate of $\hat{S}_z$ but not a sufficient one.
The situation is even more complicated for the total spin. Contrary to $\hat{S}_z$ the total spin $\hat{\mathbf{S}}^2$ is not a single-particle operator. Its expectation value is, therefore, not a trivial functional of the 1RDM. Consequently, restricting the 1RDMs to have a specific expectation value for $\hat{\mathbf{S}}^2$ is nontrivial as well since it requires the knowledge of the expectation value of $\hat{\mathbf{S}}^2$ as a functional of the 1RDM. Since this expectation value can easily be written as a functional of the 2-body reduced density matrix (2RDM), $\Gamma^{(2)}$, as \cite{L1955}
\bea
\label{ssquare}
\langle \hat{\mathbf{S}}^2 \rangle &=& -\frac{N(N-4)}{4}\\
\nonumber
&& + \sum_{\sigma_1,\sigma_2}\int d^3r_1 d^3r_2 
\Gamma^{(2)} (\br_1 \sigma_1,\br_2 \sigma_2 |\br_1 \sigma_2,\br_2 \sigma_1),
\eea
several attempts have been made to apply constraints on the 2RDM \cite{AV2005, Piris1, Piris2, Piris_Leiva_spin} which then also affect the 1RDM. The same problem arises in DFT because the density functional to calculate $\langle \hat{\mathbf{S}}^2\rangle$ is unknown. As an approximate functional, one then usually evaluates $\langle\hat{\mathbf{S}}^2\rangle$ using the Kohn-Sham Slater determinant. However, this, in general, does not yield the correct value of $\langle \hat{\mathbf{S}}^2\rangle$ of the interacting system \cite{Handy_spin_dft, spin_dft_reiher, spin_becke}.

In this work, we discuss some necessary conditions for the 1RDM of a 2-electron system to correspond to a triplet configuration. These conditions can also be derived from symmetry considerations of the triplet wave function\cite{LSH1956}. In analogy to the pure-state conditions, for systems with a triplet ground-state the exact functional would find the corresponding 1RDM in the energy minimization without applying additional constraints. As we see, the conditions are generally violated by the 1RDMs obtained from the three approximate functionals considered here, namely, the M\"uller, the BBC3, and the Power functionals, when they are not explicitly enforced. Moreover, using the necessary conditions, we show that the BBC3 and the Power functionals break the energy degeneracy between the highly polarized triplet state ($S_z=1$) and the $S_z=0$ one, which is a clear deficiency of these approximations. We also apply these conditions, in an approximate way, to systems with an even number of electrons that is larger than two. In this case, we assume that $N-2$ natural orbitals form a singlet configuration and only two active electrons form the triplet. We apply the triplet conditions to various small systems, which have a singlet ground state, to calculate the first excited triplet state. We show that, by imposing these constraints, the results for the optimal 1RDMs are closer to exact than the results obtained without imposing them. 
	In most cases, the total energies of the first excited triplet states also improve when the constraints are applied.
For the M\"uller functional we also report values of $\langle \hat{\mathbf{S}}^2\rangle$ since this functional provides an approximation for the whole 2RDM in terms of the 1RDM. For the other two functionals, BBC3 and Power, only the energy functional is available and of course expectation values of $\hat{\mathbf{S}}^2$ cannot be reported.

This paper is organized as follows: In Section \ref{sec:spinconst}, we present the necessary conditions which we consider for the triplet state of 2-electron systems and their generalization in order to be applicable to more electrons. Our results are presented in Section \ref{sec:results}, where we assess the inclusion of these constraints in RDMFT calculations as far as the optimal 1RDM and the total energy of the lowest triplet state are concerned. Finally, our conclusions are included in Section \ref{sec:conc}.

\section{Spin Constraints}\label{sec:spinconst}

Writing the 1RDM, $\gamma(\v r, \v r')$ in its spectral representation
\be
\gamma(\v r, \v r')=\sum_{j=1}^\infty n_j\varphi_j^*(\v r')\varphi_j(\v r)
\ee
with the occupation numbers $n_j$ and the natural orbitals $\varphi_j(\v r)$ one can easily express the ensemble $N$-representability conditions \cite{Coleman_1963} as
\be
\sum_{j=1}^\infty n_j=N, \quad 0\leq n_j \leq 1.
\ee
These two conditions ensure that the 1RDM corresponds to a system of $N$ fermions but not necessarily to a pure $N$-particle state. Obviously, in any practical calculation the number of natural orbitals is restricted to a finite number $M$ with $M>N$ which is a valid approximation since the occupation numbers $n_j$ fall off rapidly for $j>N$. Restricting the system not only to fermionic ensembles but actual $N$-particle states requires additional constraints which increase rapidly in number with the number of particles $N$ and the number of orbitals $M$. For small $N$ and $M$ they can be given explicitly \cite{ThesisAltunbulak, Klyachko_Math, Borland-Dennis}.

In the present work,
we allow for spin-dependent density matrices and occupation numbers, the natural orbitals, however, remain spin independent \cite{LHG2005}, i.e.,
\be
\gamma_\sigma(\v r, \v r')=\sum_{j=1}^\infty n_{j\sigma}\varphi_j^*(\v r')\varphi_j(\v r)
\ee
and
\be
\gamma(\v r, \v r')=\sum_{\sigma=\uparrow, \downarrow} \gamma_\sigma(\v r, \v r').
\ee
Note that the choice of having the same set of spatial orbitals for both spin 
channels is not related to describing density matrices with a specific $S_z$ (which can be also achieved with different sets of spin orbitals \cite{RP2011}) but,
facilitates the description of $\gamma$ with a specific expectation value of $\hat{\mathbf{S}}^2$. In  approximations which use a single Slater determinant, the introduction of spatial orbitals that are different for the two spins induces the so-called spin contamination problem (see for example Ref.~\cite{spin_contamination}). This problem is completely avoided when the orbitals are spin-independent. In analogy, in RDMFT, as we see in the present work, the assumption of spin-independent spatial orbitals leads to necessary conditions for fixing the correct expectation value of $\hat{\mathbf{S}}^2$ for two electrons which take a simple form and involve only the spin-dependent occupation numbers. From now on, with the notation $S_z$ we mean, in general, the expectation value $\langle\hat{S_z}\rangle$.

In order to describe a system with a specific $S_z$ one requires that
\be
\label{repres_spin}
\sum_{j=1}^\infty n_{j\sigma}=N_\sigma, \quad \sum_\sigma N_\sigma=N.
\ee
However, while this guarantees that the expectation value of $\hat{S}_z$ is given by $(N_\uparrow-N_\downarrow)/2$, it is only a necessary but not a sufficient condition for the 1RDM to correspond to an eigenstate of $\hat{S}_z$. For example, the state
\be
\Psi(\v r_1\sigma_1,\v r_2\sigma_2) = \frac{1}{\sqrt{2}}
\left(|1^\uparrow 2^\uparrow\rangle + |3^\downarrow 4^\downarrow\rangle\right)
\ee
is a linear combination of an $S_z=1$ and an $S_z=-1$ eigenstate. We denote the natural orbital $\varphi_1(\v r)$ being occupied with a spin-up electron as $1^\uparrow$ and the Slater determinant as $\mid \:\:\rangle$. The nonzero occupation numbers of this state are given by  
\be
n_{1\uparrow}=n_{2\uparrow}=n_{3\downarrow}=n_{4\downarrow}=\frac{1}{2},
\ee
and the sum of the occupation numbers in each spin channel is $N_\uparrow=N_\downarrow=1$. 
Thus, even when both $N_\uparrow$ and $N_\downarrow$ are fixed to integer values, the 1RDM does not need to correspond to an eigenstate of the $\hat{S}_z$ operator.  Exceptions are the maximally polarized states, i.e., for fixed $N_\uparrow=N$ and $N_\downarrow=0$ or vice versa. In these cases, one is guaranteed to find an $\hat{S}_z$ eigenstate with $S_z=\pm N/2$. 
Furthermore, for these states there exists only one $\hat{\mathbf{S}}^2$ eigenstate. Therefore, in this specific situation, enforcing a certain value for $S_z$ ensures that the 1RDM corresponds to an eigenstate of both $\hat{S}_z$ and $\hat{\mathbf{S}}^2$ with the latter having the eigenvalue $S(S+1)=(N/2)(N/2+1)$, provided one enforces pure-state $N$-representability. If pure-state $N$-representability is not enforced the calculation will generally yield an ensemble of states with $S_z=\pm N/2$. In other words, each of the states in the ensemble will be an eigenstate of $\hat{S}_z$ and $\hat{\mathbf{S}}^2$ and the expectation values of the whole ensemble will be $\pm N/2$ and $N/2(N/2+1)$, respectively. The pure-state $N$-representability can be ensured by simply transferring the known pure-state $N$-representability conditions \cite{ThesisAltunbulak, Klyachko_Math, Borland-Dennis, Carlos1, Carlos2} to the occupation numbers of the spin channel that is occupied in the system. 

For $N=2$, there are only two possible configurations for the total spin, $S=1$ or $S=0$. The fully polarized states correspond to $S_z=\pm 1$ and, as discussed above, are easy to distinguish in RDMFT from the $S_z=0$ states. The necessary and sufficient conditions for pure-state $N$-representability for $N=2$ only require a double degeneracy of the occupation numbers \cite{LSH1956}. Hence, enforcing all occupation numbers of the up (down) spin channel to be doubly degenerate yields a triplet eigenstate with $S_z=1$ ($S_z=-1$). The question remains how to distinguish between the two $S_z=0$ states, i.e., the triplet state with $S_z=0$ and the singlet state. We can construct the wave function for the triplet state with $S_z=0$ starting from the fully polarized state
\be\label{eq:trips=1}
\left|S=1, S_z=1\right\rangle=a_1|1^\uparrow 2^\uparrow\rangle+a_2|3^\uparrow 4^\uparrow\rangle+a_3|5^\uparrow 6^\uparrow\rangle \cdots\: .
\ee
Note that one needs an even number of natural orbitals $M$ since only doubly excited Slater determinants are allowed in the expansion. Including a determinant which is a single excitation of any other determinant in the expansion leads to off-diagonal terms in the 1RDM which is forbidden, as we are constructing the Slater determinants from natural orbitals. As one can see, the occupation numbers for such a state are pairwise degenerate with 
\be\label{eq:2fold}
n_{1\uparrow}=n_{2\uparrow}=|a_1|^2, n_{3\uparrow}=n_{4\uparrow}=|a_2|^2 \cdots\,,
\ee
i.e.. the pure-state constraint is satisfied. Applying $\hat{S}_-$ to the state (\ref{eq:trips=1}) we obtain
\bea\label{eq:trips=0}
|S=1, S_z=0\rangle&=&\frac{1}{\sqrt{2}} \bigl(a_1[|1^\downarrow 2^\uparrow\rangle+|1^\uparrow 2^\downarrow\rangle]\\
\nonumber
&&\quad\:\: +a_2[|3^\downarrow 4^\uparrow\rangle+|3^\uparrow 4^\downarrow\rangle]+\cdots\bigl),
\eea
where $1^\uparrow$ denotes that the natural orbital $\varphi_1$ is occupied with an up electron in the Slater determinant, and the spatial dependence of $1^\uparrow$ and $1^\downarrow$ is identical. The corresponding occupation numbers are 4-fold degenerate with 
\bea\label{eq:4fold}
&&n_{1\uparrow}=n_{1\downarrow}=n_{2\uparrow}=n_{2\downarrow}=|a_1|^2/2,\\\nonumber
&&n_{3\uparrow}=n_{3\downarrow}=n_{4\uparrow}=n_{4\downarrow}=|a_2|^2/2,\\
\nonumber
&&\cdots
\eea

We also see from Eqs.\ (\ref{eq:trips=1}) and (\ref{eq:trips=0}) that the spatial parts of the natural orbitals of the two spin channels are the same since the spin operator only acts on the spin parts. Corresponding to the triplet $S_z=0$ state there also exists a singlet state  
\bea\label{eq:sing}
|S=0, S_z=0\rangle &=&\frac{1}{\sqrt{2}}  \bigl(a_1[|1^\downarrow 2^\uparrow\rangle-|1^\uparrow 2^\downarrow\rangle]\\
\nonumber
&&\quad\:\: +a_2[|3^\downarrow 4^\uparrow\rangle-|3^\uparrow 4^\downarrow\rangle]+\cdots\bigl)
\eea
which also has the occupation numbers given by Eq.\ (\ref{eq:4fold}). In other words, the 4-fold degeneracy of the occupation numbers is a necessary but not sufficient condition for the 1RDM to belong to a triplet state with $S_z=0$. As a result, contrary to the fully polarized states, enforcing a 4-fold degeneracy on the occupation numbers might still yield a singlet state rather than the $S_z=0$ triplet state. However, this is not the most general singlet state. One can also construct a singlet as
\be
|S=0, S_z=0\rangle=c_1|1^\downarrow 1^\uparrow\rangle
+c_2|2^\downarrow 2^\uparrow\rangle+\cdots
\label{eq:gensing}
\ee
which leads to occupation numbers that are doubly degenerate only. The double degeneracy in the occupation numbers of the singlet 1RDM and the spatial part of the triplet 1RDM were derived from symmetry considerations of the corresponding wave functions by L\"owdin and Shull \cite{LSH1956}.

Running a RDMFT calculation with $N_\uparrow=N_\downarrow=1$ without enforcing extra constraints we typically find the double degeneracy of the occupations that corresponds to the general singlet configuration, Eq.\ (\ref{eq:gensing}). This is true not only for approximate functionals but also for the exact one which is known for $N=2$ \cite{LSH1956}. This is not surprising since the occupation numbers have more variational freedom than for the states (\ref{eq:trips=0}) and (\ref{eq:sing}) where a 4-fold degeneracy is required. Even a linear combination of the general singlet state (\ref{eq:gensing}) and the triplet state (\ref{eq:trips=0}) yields some occupation numbers which are 4-fold degenerate. For example, the state
\be\label{eq:lincomb}
c_1|1^\downarrow 1^\uparrow\rangle +
c_2\left(|2^\downarrow 3^\uparrow\rangle+|2^\uparrow 3^\downarrow\rangle\right),
\ee
where the first part is a singlet state while the second one is a $S_z=0$ triplet state, corresponds to occupation numbers
\bea\label{eq:lincombocc1}
n_{1\uparrow}&=&n_{1\downarrow}=|c_1|^2,\\
\label{eq:lincombocc2}
n_{2\uparrow}&=&n_{2\downarrow}=n_{3\uparrow}=n_{3\downarrow}=|c_2|^2.
\eea
In other words, the occupation numbers coming from the triplet part are again 4-fold degenerate. Note that due to the fact that we are expanding in natural orbitals, an orbital from the singlet part of Eq.\ (\ref{eq:lincomb}) cannot be used again in the triplet part since this would introduce determinants which are single excitations of each other.

Energetically, unless the Hamiltonian contains a magnetic field or any other spin-specific terms, the two states (\ref{eq:trips=1}) and (\ref{eq:trips=0}) are degenerate. Hence, for calculating the triplet energy it should be irrelevant which state is calculated. However,  many RDMFT functionals do not satisfy this degeneracy. An exception is the M\"uller functional for which one can show that the states (\ref{eq:trips=1}) and (\ref{eq:trips=0}) have the same energy (see Appendix \ref{app:mueller}). In those cases where the degeneracy is broken, one can calculate the $S_z=0$ triplet state by enforcing the 4-fold degeneracy of the occupation numbers. This prevents the minimization from finding the general singlet state (\ref{eq:gensing}) which is lower in energy. The lowest singlet state with 4-fold degeneracy (\ref{eq:sing}) has a higher energy than the corresponding triplet state (\ref{eq:trips=0}) \cite{Atkins}.

For more than two electrons, one often encounters cases where only the two outer electrons are important for describing the correct physics. Within any multiconfiguration wave function approach, this corresponds to working with only two active electrons. 
For example, for four electrons, one writes the wave function as
\bea\nonumber
|S=1, S_z=1\rangle &=& a_1|1^\uparrow1^\downarrow 2^\uparrow 3^\uparrow\rangle+a_2|1^\uparrow1^\downarrow 4^\uparrow 5^\uparrow\rangle\\
\label{eq:M=1_pinned}
&+& a_3|1^\uparrow1^\downarrow 6^\uparrow 7^\uparrow\rangle + \ldots
\eea
which leads to the following occupation numbers
\bea
\label{eq:occM=1_pinned}
\nonumber
n_{1\uparrow}&=& n_{1\downarrow}=1,\\
\nonumber
n_{2\uparrow}=n_{3\uparrow},&& n_{4\uparrow}=n_{5\uparrow},\\
\nonumber
n_{6\uparrow}=n_{7\uparrow},&& \ldots\: .\\
\eea
Acting with $\hat{S}_{-}$ on the state of Eq.~(\ref{eq:M=1_pinned}) we obtain 
\bea
\nonumber
|S=1,S_z=0\rangle &=& \frac{a_1}{\sqrt{2}}\left(|1^\uparrow 1^\downarrow 2^\uparrow 3^\downarrow \rangle+|1^\uparrow 1^\downarrow 2^\downarrow 3^\uparrow\rangle\right)\\
\nonumber
&+&\frac{a_2}{\sqrt{2}}\left(|1^\uparrow 1^\downarrow 4^\uparrow 5^\downarrow\rangle + |1^\uparrow 1^\downarrow 4^\downarrow 5^\uparrow\rangle\right)\\
\nonumber
&+&\frac{a_3}{\sqrt{2}}\left(|1^\uparrow 1^\downarrow 6^\uparrow 7^\downarrow\rangle + |1^\uparrow 1^\downarrow 6^\downarrow 7^\uparrow\rangle\right)\\
&+&\ldots
\label{eq:M=0_pinned}
\eea
with occupation numbers
\bea
\label{eq:occM=0_pinned}
\nonumber
n_{1\uparrow}&=& n_{1\downarrow}=1,\\
\nonumber
n_{2\uparrow}=n_{2\downarrow}&=&n_{3\uparrow}=n_{3\downarrow},\\
\nonumber
n_{4\uparrow}=n_{4\downarrow}&=&n_{5\uparrow}=n_{5\downarrow},\\
\nonumber
n_{6\uparrow}=n_{6\downarrow}&=&n_{7\uparrow}=n_{7\downarrow},\\
\vdots && \vdots \: .
\eea
However, for the $|S=1,S_{z}=1\rangle$, the choice of forcing all the inner orbitals to have occupation numbers equal to one is very restrictive and leaves no variational freedom in the spin-down channel. There are cases, as we see later, where imposing the corresponding constraints in an energy minimization leads to an overestimation of the energy of the triplet state. As an alternative, we suggest to assume that the inner orbitals are equally, but not necessarily fully, occupied in the two spin channels, such that these orbitals give a $S=0$ contribution to the total spin. Hence, we enforce the following conditions on the occupation numbers for $S_z=1$  
\bea
\label{eq:occM=1}
\nonumber
n_{1\uparrow}&=& n_{1\downarrow},\\
\nonumber
n_{2\uparrow}=n_{3\uparrow},&& n_{4\uparrow}=n_{5\uparrow},\\
\nonumber
n_{6\uparrow}=n_{7\uparrow}\\
\vdots && \vdots\:,
\eea
and leave the occupation numbers of the down channel, except for $n_{1\downarrow}$, unconstrained.
The conditions for four electrons of Eqs.~(\ref{eq:occM=1_pinned}), (\ref{eq:occM=0_pinned}), and (\ref{eq:occM=1}) can be extended to any even number of electrons. One can then apply them to calculate, for example, approximate total energies for the lowest triplet state. We expect these total energies to be correct for the lowest triplet state because the assumption that this triplet state is built entirely by the two outer electrons is a good approximation in this case. 
We would like to point out that with the constraint (\ref{eq:occM=1_pinned}), the system is effectively reduced to a 2-particle system with $S_z=1$. Hence, the double degeneracy of the fractional occupation numbers is a necessary and sufficient condition to find expectation values $S=1$ and $S_z=1$. Using instead the constraint (\ref{eq:occM=1}), the calculated 1RDM will not correspond exactly to the correct expectation value of $\hat{\mathbf{S}}^2$ due to the additional freedom especially in the outer occupation numbers of the spin down channel. Nevertheless, this constraint will improve the description of a triplet state in general compared with no constraint at all. The effect of the constraints in the description of triplets can be tested for the M\"uller functional which offers an ansatz for the full 2RDM and the expectation value of $\hat{\mathbf{S}}^2$ can be calculated.

We note that enforcing the pinning of some occupation numbers to one, which is generally an approximation, reduces dramatically the number of pure-state $N$-representability conditions and makes it easier to apply them in practical implementations. As the number of electrons and the dimension of the Hilbert space increase, the number of the exact generalized Pauli constraints explodes and their consideration in the minimization is extremely difficult. Pinning an occupation number means that it has no influence in the question of pure-state or ensemble $N$-representability. From Eqs.\ (\ref{eq:M=1_pinned})-(\ref{eq:occM=0_pinned}), we can see that the orbitals which correspond to pinned occupation numbers appear in every Slater determinant in a pure state. For an ensemble, such an  orbital would appear in every Slater determinant of every term that contributes to the ensemble. Consequently, for those systems where pinning the occupation numbers for all but two electrons is a valid approximation, we could simply consider the generalized Pauli constraints for two electrons and for those occupation numbers which are not pinned. For a 2-electron triplet with $S_z=\pm 1$ the double degeneracy constraint (\ref{eq:2fold}) coincides with the generalized Pauli constraint. For the triplet with $S_z=0$, the 4-fold degeneracy constraint (\ref{eq:4fold}) is stricter but fulfills the generalized Pauli constraints.

So far, we have discussed constraints on the triplet states. In order to calculate the excitation energy from a singlet ground state to the first excited triplet state, we also need to calculate the energy of the singlet. Since many approximations were derived aiming at the correct description of singlet ground states, we expect to obtain quantitatively correct results for the singlet ground-state energies by just enforcing ${S}_z=0$ despite the fact that we cannot exclude that our density matrix might be contaminated by contributions from states with total spin $S$ larger than zero. Note that for systems with an even number of electrons, a nondegenerate ground state and a Hamiltonian which has time-reversal symmetry, by enforcing ${S}_z=0$ setting the constraint $n_{j\uparrow}=n_{j\downarrow}$, we also satisfy the generalized Pauli constraints \cite{S1966}.

Let us point out that even when the generalized Pauli constraints are satisfied this does not mean that the 1RDM that we have corresponds necessarily to a pure state, it could also correspond to an ensemble. However, by satisfying these constraints we exclude the case that a given 1RDM corresponds to ensembles only and cannot correspond to a pure state.

\section{Results\label{sec:results}}

We now apply the constraints discussed in the last section to 
the energy minimization in RDMFT calculations. We employed three different  approximations for the total energy within RDMFT, namely, the M\"uller, the BBC3, and the Power functionals \cite{M1984, GPB2005, SDLG2008}. We first compare the 1RDMs obtained from the constrained calculations to those obtained without these constraints and the ``exact'' 1RDM from a MCSCF calculation. The inclusion of the extra constraints on the occupation numbers adds only an insignificant amount of computational cost since the bottleneck of the RDMFT minimization is the optimization of the natural orbitals. We refer to the calculations without the constraints discussed in this work, related to $\langle\hat{\mathbf{S}}^2\rangle$, as minimizations without constraint. Despite this name, we still impose the constraint (\ref{repres_spin}) which fixes the expectation value of $\hat{S_z}$ and the correct number of electrons in all our calculations.

As test systems, we considered helium, H$_2$, Be, BH, H$_2$O and Mg for which, due to their small size, the MCSCF calculations are feasible. For  helium, H$_2$, Be, and BH, we used the cc-pVTZ basis set, and the energy minimization for all methods was performed using $10$, $14$, $35$, and $24$ natural orbitals, respectively. For H$_2$O and Mg we used the cc-pVDZ with $20$ and $17$ natural orbitals, respectively. In all cases, we used less natural orbitals than the basis sets would allow, since, for small systems, we obtained very small occupation numbers which cause numerical problems in the convergence of the MCSCF calculations. In addition, for larger systems the demands in memory become prohibitive for the MCSCF calculation because we want to compare to the exact 1RDM and, therefore, cannot pin occupations to one. The MCSCF triplet as well as the one- and 2-electron integral calculations were performed using the Gamess US code \cite{GAMESS}.  The RDMFT calculations were performed with the HIPPO computer code \cite{code}.

\subsection{Quality of 1-RDMs Using $\langle \hat{\mathbf{S}}^2 \rangle$ Constraints}

As discussed in the previous section, a 2-electron wave function with $S_z=1$ expressed in terms of natural orbitals has the form (\ref{eq:trips=1}), which results in the restrictions (\ref{eq:2fold}). Thus, we tested if these exact conditions are satisfied by the functionals considered here, for the helium atom and the hydrogen molecule at different internuclear separations by calculating the ground state for $N_\uparrow=2$ and $N_\downarrow=0$. We find that the conditions are violated in all cases, i.e.. the occupations are not pairwise equal but show differences of up to $0.09$ within the pairs. The conditions are, as expected, satisfied by the MCSCF calculations and, as can be shown analytically, by the exact 2-electron RDMFT functional (LSH) \cite{LSH1956}. We then perform the RDMFT calculations using Eq.\ (\ref{eq:2fold}) as an additional constraint during the optimization of the occupation numbers.

We also perform RDMFT minimizations for larger systems with an even number of particles with the approximate constraint (\ref{eq:occM=1_pinned}), where we assume that the triplet is formed from the two outermost electrons and the inner orbitals have occupations pinned to one and with the constraint (\ref{eq:occM=1}), where we allow the inner occupations to be less than one. We compare the occupation numbers from these RDMFT minimizations, with and without enforcing the constraints, with occupation numbers from ``exact'' MCSCF calculations to check whether the constraints help to get a density matrix closer to the exact one. 

As the constraints concern the occupation numbers, an important criterion for the quality of the calculated 1RDMs is the square difference
\be\label{eq:delta}
\Delta=\frac{1}{N}\sum_{j=1}^M \left(\sum_{\sigma=\uparrow,\downarrow}\left(n_{j\sigma}^\mathrm{RDMFT}-n_{j\sigma}^\mathrm{MCSCF}\right)\right)^2
\ee
of the obtained RDMFT occupations from the exact ones, which we show in Table \ref{tab:dev_occ}. In Eq.\ (\ref{eq:delta}), $M$ denotes the number of natural orbitals included in the calculation and $N$ the total number of electrons. We show results for $\Delta$ without imposing the additional constraints (w/o) and from calculations with the spin constraint (\ref{eq:2fold}) for two electrons (cons.) in the top half of the Table. For more than two electrons, we impose the constraints of Eq.~(\ref{eq:occM=1_pinned}) (cons.\ pin.) and (\ref{eq:occM=1}) (cons.), and the results are shown in the bottom half. For the 2-electron systems, imposing the exact constraint (\ref{eq:2fold}), in calculations, with the approximate 1RDM functionals we adopted, results in optimal occupation numbers closer to the exact ones. Moreover, for more than two electrons, both approximate constraints, Eqs.\ (\ref{eq:occM=1_pinned}) and (\ref{eq:occM=1}), improve significantly the 1RDM as the occupation numbers are much closer to the exact ones than the occupations from the energy minimization without additional constraints.
We should emphasize that, although the constraint (\ref{eq:occM=1}) is expected to lead to a larger deviation from the correct $\langle \hat{\mathbf{S}}^2\rangle$ than the constraint (\ref{eq:occM=1_pinned}), it leads to occupations closer to the exact ones.

\begin{table*}
\setlength{\tabcolsep}{1.0pt}
\begin{tabular}{|l|ccc|ccc|ccc|} \hline\hline
& \multicolumn{3}{c|}{M\"uller} & \multicolumn{3}{c|}{BBC3} & \multicolumn{3}{c|}{Power}   \\
& w/o  &   \multicolumn{2}{c|}{cons.} & w/o &  \multicolumn{2}{c|}{cons.}  & w/o & \multicolumn{2}{c|}{cons.}  \\[0.1ex]\hline
He        & 0.001505  & \multicolumn{2}{c|}{0.000063} & 0.000108 & \multicolumn{2}{c|}{0.000014} & 0.000109 & \multicolumn{2}{c|}{0.000005}\\ 
H$_2$ (1.4 au) & 0.019184  &	\multicolumn{2}{c|}{0.000855} & 0.000743 & \multicolumn{2}{c|}{0.000006} & 0.002157 & \multicolumn{2}{c|}{0.000060} \\
H$_2$ (2.5 au) & 0.004083  &	\multicolumn{2}{c|}{0.000853} & 0.000312 & \multicolumn{2}{c|}{0.000064} & 0.000377 & \multicolumn{2}{c|}{0.000063} \\
H$_2$ (5.0 au) & 0.001690  &	\multicolumn{2}{c|}{0.000808} & 0.000294 & \multicolumn{2}{c|}{0.000089} & 0.000161 & \multicolumn{2}{c|}{0.000075} \\\hline
average $\Delta$ & 0.00662   &  \multicolumn{2}{c|}{0.00064} & 0.00036   & \multicolumn{2}{c|}{0.00004} & 0.00070  & \multicolumn{2}{c|}{0.00005} \\\hline\hline
& w/o  &   cons. pin. & cons. & w/o &  cons. pin    &  cons.  & w/o & cons. pin.& cons.  \\[0.1ex] \hline
Be        & 0.207842  &	0.090246 & 0.068044 & 0.001813 & 0.000302 & 0.000166  & 0.164014 & 0.022162 & 0.010776  \\
BH        & 0.081441  &	0.023143 &0.0091545& 0.028337& 0.000731 & 0.000274 & 0.078891 & 0.004478& 0.001066 \\ 
H$_2$O    & 0.034401  & 0.003178 &0.0077090& 0.001668 & 0.000476 &0.000152 & 0.011851 & 0.000708 & 0.000877  \\
Mg        & 0.053385  &	0.034422 &0.0162205& 0.011861 &	0.001604 &0.000128 & 0.053019 &	0.003068 & 0.003617\\\hline 
average $\Delta$& 0.09427 &0.03775 &0.02528&0.01092&0.00078&0.00018&0.07694&0.00760   &	0.00408\\
 \hline\hline
\end{tabular}
\caption{\label{tab:dev_occ}{\bf For the $S_z=1$ State, deviation of the calculated occupation numbers from the exact.} For two-electron systems (top): Deviation $\Delta$ (see Eq.\ (\ref{eq:delta})) of the occupation numbers from the exact occupations (MCSCF) calculated with different RDMFT functionals, without (w/o) enforcing the additional exact spin constraint (\ref{eq:2fold}) and with the constraint (cons.). For systems with more than two electrons (bottom): The same deviation without any constraint (w/o), with the constraint (\ref{eq:occM=1_pinned}) (cons. pin.), or using Eq.\ (\ref{eq:occM=1}) (cons.). For each system we used the same number of natural orbitals and the same basis set for the RDMFT and MCSCF calculations.}
\end{table*}
\begin{table*}
\setlength{\tabcolsep}{1.0pt}
\begin{tabular}{|l|ccc|ccc|ccc|c|} \hline\hline
& \multicolumn{3}{c|}{M\"uller} & \multicolumn{3}{c|}{BBC3} & \multicolumn{3}{c|}{Power}&MCSCF   \\
& w/o  &   \multicolumn{2}{c|}{cons.} & w/o & \multicolumn{2}{c|}{cons.}  & w/o & \multicolumn{2}{c|}{cons.} &  \\[0.1ex] \hline
He     &  4.9$\cdot 10^{-2}$ & \multicolumn{2}{c|}{1.3$\cdot 10^{-2}$}  & 1.3$\cdot 10^{-2}$ & \multicolumn{2}{c|}{6.3$\cdot 10^{-3}$} &1.3$\cdot 10^{-2}$  & \multicolumn{2}{c|}{3.8$\cdot 10^{-3}$} &1.9$\cdot 10^{-4}$ \\
H$_2$ (1.4 au) &  1.0$\cdot 10^{-1}$ & \multicolumn{2}{c|}{5.5$\cdot 10^{-2}$}  &3.8$\cdot 10^{-2}$  & \multicolumn{2}{c|}{1.7$\cdot 10^{-2}$}  & 6.1$\cdot 10^{-2}$ & \multicolumn{2}{c|}{1.9$\cdot 10^{-2}$} & 5.7$\cdot 10^{-3}$ \\
H$_2$ (2.5 au) &  8.5$\cdot 10^{-2}$ & \multicolumn{2}{c|}{5.4$\cdot 10^{-2}$}  &2.5$\cdot 10^{-2}$  & \multicolumn{2}{c|}{1.7$\cdot 10^{-2}$}  & 2.7$\cdot 10^{-2}$ & \multicolumn{2}{c|}{1.7$\cdot 10^{-2}$} & 3.0$\cdot 10^{-3}$ \\
H$_2$ (5.0 au) &  5.4$\cdot 10^{-2}$ & \multicolumn{2}{c|}{5.1$\cdot 10^{-2}$} &2.3$\cdot 10^{-2}$ & \multicolumn{2}{c|}{1.7$\cdot 10^{-2}$}  & 1.7$\cdot 10^{-2}$ & \multicolumn{2}{c|}{1.5$\cdot 10^{-4}$}  & 2.5$\cdot 10^{-4}$\\\hline
average $\Delta_w$ & 7.0$\cdot 10^{-2}$	&\multicolumn{2}{c|}{4.1$\cdot10^{-2}$}&2.3$\cdot 10^{-2}$&\multicolumn{2}{c|}{1.2$\cdot10^{-2}$}&2.7$\cdot10^{-2}$&\multicolumn{2}{c|}{1.1$\cdot 10^{-2}$} &  -\\\hline \hline
& w/o  &   cons. pin. & cons. & w/0 &  cons. pin    &  cons.  & w/o & cons. pin.& cons.&  \\[0.1ex] \hline
Be               &  6.8$\cdot 10^{-1}$ & 6.8$\cdot 10^{-1}$ &5.9$\cdot 10^{-1}$& 8.2$\cdot 10^{-2}$  & 5.5$\cdot 10^{-2}$  &4.3$\cdot 10^{-2}$& 1.0 & 3.3$\cdot 10^{-1}$ &2.3$\cdot 10^{-1}$& 1.5$\cdot 10^{-2}$\\
BH               &  1.1& 4.5$\cdot 10^{-1}$ &3.8$\cdot 10^{-1}$& 3.4$\cdot 10^{-1}$  & 7.2$\cdot 10^{-2}$  &5.5$\cdot 10^{-2}$& 1.0 & 2.0$\cdot 10^{-1}$ & 1.3$\cdot 10^{-1}$&7.7$\cdot 10^{-2}$ \\
H$_2$O           &  4.1$\cdot 10^{-1}$ & 2.3$\cdot 10^{-1}$&2.0$\cdot 10^{-1}$& 1.1$\cdot 10^{-1}$  & 8.6$\cdot 10^{-2}$  &6.7$\cdot 10^{-2}$ & 2.3$\cdot 10^{-1}$ & 1.1$\cdot 10^{-1}$&7.9$\cdot 10^{-2}$ & 8.1$\cdot 10^{-2}$ \\
Mg   &  1.0 &7.1$\cdot 10^{-1}$ &5.0$\cdot 10^{-1}$& 3.0$\cdot 10^{-1}$  &1.6$\cdot 10^{-1}$   &5.6$\cdot 10^{-2}$ & 1.0 &2.3$\cdot 10^{-1}$& 2.3$\cdot 10^{-1}$ & 1.3$\cdot 10^{-2}$\\\hline
average $\Delta_w$& 7.5$\cdot10^{-1}$& 4.7$\cdot 10^{-1}$&3.3$\cdot 10^{-1}$&1.6$\cdot 10^{-1}$&5.0$\cdot10^{-2}$&1.5$\cdot10^{-2}$&7.7$\cdot10^{-1}$&7.2$\cdot10^{-1}$&1.2$\cdot10^{-1}$&-\\
\hline\hline
\end{tabular}
\caption{\label{tab:weak_occ} {\bf Same as Table \ref{tab:dev_occ} but for the sum of the occupation numbers of the weakly occupied orbitals, $w$ (see Eq.\ (\ref{eq:wcorr})).} Here $\Delta_w$ denotes the  absolute deviation from the ``exact'' MCSCF results which is then averaged over all systems in each part of the table.}
\end{table*}

Correlations in RDMFT are manifested by fractional occupation numbers. A measure for the correlation is the total electronic charge of ``weakly'' occupied orbitals, i.e., those with occupations smaller than $1/2$, defined as
\be\label{eq:wcorr}
w=\sum_{n_{j\sigma} <\frac{1}{2}} n_{j\sigma}.
\ee
We again compare to the results from a MCSCF calculation using 
\begin{eqnarray}
\Delta_w=\left|w-w^{MCSCF}\right|
\end{eqnarray}
which is then averaged over all the systems considered.
A known deficiency of many approximate functionals, which can also be seen in Table \ref{tab:weak_occ}, is that they typically overestimate $w$. For 2-electron systems, imposing the exact constraint (\ref{eq:2fold}) for $S=1$ and $S_z=1$, lowers $w$ to values closer to the exact ones. For systems with more than two electrons, both the constraints (\ref{eq:occM=1_pinned}) and (\ref{eq:occM=1}) reduce $w$ to values closer to the ``exact" MCSCF result.
Results with the constraint (\ref{eq:occM=1}), i.e., without pinning the inner occupation numbers, are the closest to MCSCF.

\begin{figure}
\includegraphics[width=0.4\textwidth, clip]{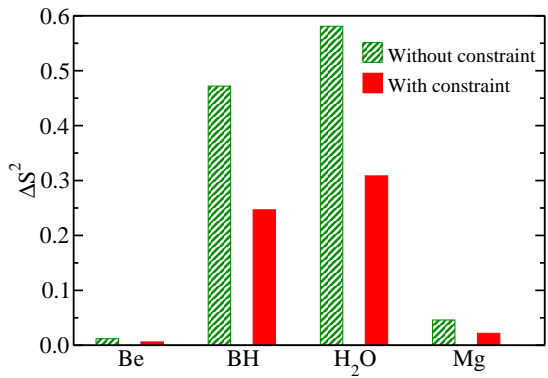}
\caption{\label{fig:spin_violation} Difference between  $\langle\hat{\mathbf{S}}^2 \rangle$ calculated from Eq.\ (\ref{eq:Stot_Mueller}) with the M\"uller functional and the exact $\langle\hat{\mathbf{S}}^2 \rangle=2$ for a triplet state. The triplet is calculated without any constraints for $\hat{\mathbf{S}}^2$ (green shaded) and with the constraint Eq.\ (\ref{eq:occM=1}) (red full).}
\end{figure}

So far, we have assessed the quality of the optimal 1RDMs from our calculations by comparing them to the results from MCSCF calculations. However, the goal was to calculate 1RDMs with a specific expectation value for the total spin which requires a functional of $\langle \hat{\mathbf{S}}^2\rangle$ in terms of the 1RDM. For the M\"uller functional, the energy functional was derived using an ansatz for the second-order reduced density matrix $\Gamma^{(2)}$ in terms of the 1RDM \cite{M1984}. Using this ansatz for $\Gamma^{(2)}$ we can calculate $\langle \hat{\mathbf{S}}^2\rangle$ from Eq.\ (\ref{ssquare}). The resulting expression reads
\begin{eqnarray}\nonumber
 \langle \hat{\mathbf{S}}^2 \rangle_{\mathrm{M\ddot{u}ller}}& = &\frac{(N_\uparrow-N_\downarrow)^2}{4} 
+ (N_\uparrow + N_\downarrow) \\
&&-\frac{1}{2} \sum_{j=1}^{\infty} \sum_{\sigma\sigma^\prime = \uparrow,\downarrow} 
\sqrt{(n_{j\sigma}\,n_{j\sigma^\prime})}.  
\label{eq:Stot_Mueller}
\end{eqnarray}
The correct expectation value of the triplet state is $\langle \hat{\mathbf{S}}^2 \rangle=2 $. Thus, we calculate the difference
\begin{eqnarray}
\Delta \mathbf{S}^2=2-\langle \hat{\mathbf{S}}^2 \rangle_{\mathrm{M\ddot{u}ller}}
\end{eqnarray}
to examine whether imposing the constraint (\ref{eq:occM=1}) improves 
$\langle \hat{\mathbf{S}}^2 \rangle$ or not. The results we obtained are shown in Fig.\ \ref{fig:spin_violation}. It is apparent that in all cases the
considered approximate constraint improves the values of $\langle \hat{\mathbf{S}}^2 \rangle$ obtained with Eq.~(\ref{eq:Stot_Mueller}). Note that with the constraint (\ref{eq:occM=1_pinned}), where we pin the inner occupation numbers to one, and for 2-electron systems, where 
Eq.\ (\ref{eq:2fold}) is the exact constraint, $\Delta\mathbf{S}^2$
is zero and therefore not included in Fig.\ \ref{fig:spin_violation}.

Let us point out that the M\"uller ansatz for $\Gamma^{(2)}$ is not exact, therefore, the calculated value $\langle \hat{\mathbf{S}}^2 \rangle_{\mathrm{M\ddot{u}ller}}$ is also not exact. However, the value is consistent with the energy functional which was used in the minimization procedure. Unfortunately, for the other approximations that we employed, the  functionals for $\langle \hat{\mathbf{S}}^2 \rangle$ that are consistent with the energy functionals are not available.

\subsection{{Energy of Excited Triplet States}\label{sec:sing-trip}}

In this subsection, we discuss the effect of imposing the constraints for $\langle \hat{\mathbf{S}}^2\rangle$ on the total energies of excited triplet states. By imposing the constraint (\ref{eq:2fold}) we calculate the lowest lying triplet energy of 2-electron systems with $S_z=1$ and compare it with the energy that we get using the constraint (\ref{eq:4fold}) for the triplet with $S_z=0$. As we show in Appendix \ref{app:mueller}, the M\"uller functional respects the degeneracy between the $S_z=0$ triplet state and the fully polarized triplet states. This is not the case for the other approximations we employed in this work.

\begin{figure}
\includegraphics[width=0.45\textwidth, clip]{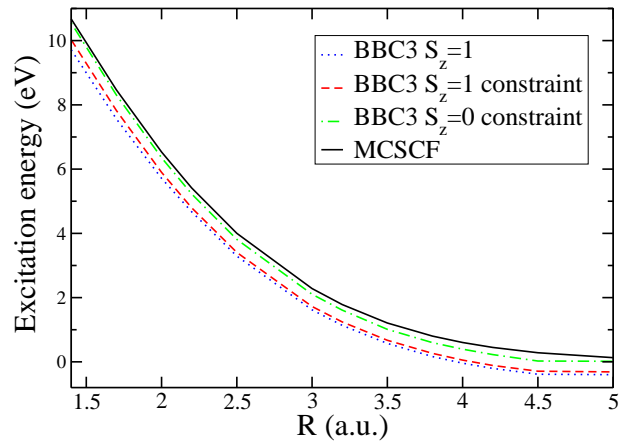}
\caption{\label{fig:h2diss} Excitation energy, ground-state singlet to the lowest triplet, of the hydrogen molecule as a function of the internuclear distance, using the BBC3 functional. The triplet is calculated either by just imposing $S_z=1$ with no constraint for  $\langle \hat{\mathbf{S}}^2\rangle$ or by additionally imposing the constraint for $S=1$, $S_z=1$ (Eq. (\ref{eq:2fold})) or for $S=1$, $S_z=0$ (Eq. (\ref{eq:4fold})).}
\end{figure}

In Fig.\ \ref{fig:h2diss}, we plot the energy difference between the first excited triplet and the ground-state singlet of the H$_2$ molecule, i.e., the first singlet to triplet excitation energy, as a function of the internuclear distance, using the BBC3 functional. The unconstrained calculation for $S_z=1$ agrees only qualitatively with the MCSCF results. Enforcing the constraint (\ref{eq:2fold}) for $S=1, S_z=1$ slightly improves the excitation energies. The best results are obtained by enforcing the constraint (\ref{eq:4fold}) for $S=1, S_z=0$. The difference in the two constrained calculations arises from the fact that the degeneracy of $S=1, S_z=1$ and $S=1, S_z=0$ is broken by the functional, although it should not since the Hamiltonian is spin-independent. In Table \ref{tab:sing-tripl} (top), we show the lowest triplet total energies of the helium atom and the hydrogen molecule at two different interatomic distances, the equilibrium distance, $R$=1.4~au and a larger one, $R$=2.5~au, for all functionals that we used. For the M\"uller functional, in all systems considered, the constraints for $\langle \hat{\mathbf{S}}^2\rangle$ improve the corresponding energies compared to the unconstrained $S_z=1$ calculation. For the BBC3 functional, the $S=1$, $S_z=1$ and  $S=1$, $S_z=0$ constraints give different results. Although both improve the triplet energies the second performs better. The Power functional also breaks the energy degeneracy but only the $S=1$, $S_z=1$ constraint improves the total energies of the triplets while the  $S=1$, $S_z=0$ deteriorates them.
As a measure for the quality of the functional in calculating triplet energies we include the average, absolute, relative deviation from the MCSCF energies, i.e.,
\be
\label{eq:diff}
\delta = \frac{1}{N_{\rm sys}} \sum_i \frac{|E_i-E_i^{\rm MCSCF}|}{|E_{i}|}\,,
\ee
where $E_i$ is the RDMFT energy of system $i$, $E_i^{\rm MCSCF}$ the corresponding MCSCF energy and $N_{\rm sys}$ the number of cases. This quantity is included in Table \ref{tab:sing-tripl}. In the same Table, we also include $\delta_{\rm ex}$, defined similarly to $\delta$ in Eq.~(\ref{eq:diff}), for the energy differences between the ground-state singlets and the first excited triplet states. In the same Table, for completeness, we also include the total energies of the ground-state singlets. As one can see, the errors for the M\"uller and the BBC3 functionals in calculating these excitations are mainly due to the bad description of the triplets, as the total energies of the singlets are very accurate. In all cases, the conditions for the triplet improve the singlet-to-lowest-triplet excitation energies although with the Power functional and the $S=1$, $S_z=0$ constraint this is only achieved due to a cancelation of errors.
\begin{table*}
\setlength{\tabcolsep}{2.0pt}
{\footnotesize\begin{tabular}{|ll|c|c|c|c|} \hline
 & & M\"uller  & BBC3 & Power  & MCSCF \\ \hline\hline
He & S=1 $S_z=1 $  w/o   & -1.9809 & -1.9623 & -1.9475 & -1.9364 \\
   & S=1 $S_z=1 $  cons  & -1.9645 & -1.9565 & -1.9428 &  \\
   & S=1 $S_z=0 $  cons  & -1.9645 & -1.9515 & -1.8501 &  \\
   & S=0                 & -2.9062 & -2.8971 & -2.9022 & -2.8989 \\\hline
H$_2$ 1.4 au&  S=1 $S_z=1$ w/o  &  -0.8489  & -0.8131 & -0.7972 & -0.7794 \\
              &  S=1 $S_z=1$ cons	&  -0.8226  & -0.8017 & -0.7881 &  \\
              &  S=1 $S_z=0$ cons &  -0.8226  & -0.7824 & -0.7201 &  \\
              &  S=0              &  -1.1870  & -1.1701 & -1.1464 & -1.1716 \\\hline
H$_2$ 2.5 au&  S=1 $S_z=1$  w/o &  -0.9968  & -0.9726 & -0.9575 & -0.9445 \\
              &  S=1 $S_z=1$ cons &  -0.9873  & -0.9684 & -0.9544 &  \\
              &  S=1 $S_z=0$ cons &  -0.9873  & -0.9534 & -0.8857 &  \\
              &  S=0              &  -1.1185  & -1.0936 & -1.0614 & -1.0915 \\  \hline   
$\delta$ & S=1 $S_z=1$ w/o  & 0.056 & 0.029 & 0.014 & \\
         & S=1 $S_z=1$ cons & 0.038 & 0.021 & 0.008 & \\
         & S=1 $S_z=0$ cons & 0.038 & 0.007 & 0.061 & \\
         & S=0              & 0.013 & 0.001 & 0.017 & \\\hline
$\delta_{\rm ex}$ & S=1 $S_z=1$ w/o  & 0.12 & 0.10 & 0.15 & \\
                  & S=1 $S_z=1$ cons & 0.07 & 0.08 & 0.13 & \\
                  & S=1 $S_z=0$ cons  & 0.07 & 0.03 & 0.11 & \\
\hline
\multicolumn{6}{c}{ } \\
\hline
Be &S=1 $S_z=1$ w/o         & -14.6966 & -14.5685 & -14.5853 & -14.5327\\
   &S=1 $S_z=1$ cons. pin.  & -14.6513 & -14.5544 & -14.5514 & \\
   &S=1 $S_z=1$ cons.       & -14.6548 & -14.5582 & -14.5530 & \\
   &S=1 $S_z=0$ cons.       & -14.6958 & -14.5524 & -14.5225 & \\
   &S=0                     & -14.7471 & -14.6491 & -14.6170 & -14.6331 \\\hline
BH &S=1 $S_z=1$ w/o         & -25.3748 & -25.2098 & -25.2058 & -25.1901 \\
   &S=1 $S_z=1$ cons. pin.  & -25.2710 & -25.1811 & -25.1720 & \\
   &S=1 $S_z=1$ cons.       & -25.3109 & -25.1997 & -25.1617 & \\
   &S=1 $S_z=0$ cons.       & -24.4076 & -25.1793 & -25.1824 & \\
   &S=0                     & -25.4504 & -25.2512 & -25.2350 & -25.2385 \\\hline
H$_2$O &S=1 $S_z=1$ w/o         & -76.1195 & -75.9048 & -75.8544 & -75.8749  \\   
       &S=1 $S_z=1$ cons. pin.  & -75.8729 & -75.8061 & -75.7558 & \\   
       &S=1 $S_z=1$ cons.       & -76.0106 & -75.8704 & -75.8015 & \\   
       &S=1 $S_z=0$ cons.       & -76.1468 & -75.8616 & -75.8586 & \\   
       &S=0                     & -76.2046 & -76.3443 & -76.1097   & -76.1732            \\\hline
Mg &S=1 $S_z=1$ w/o         & -199.6822 & -199.5835 & -199.6077 & -199.5500 \\
   &S=1 $S_z=1$ cons. pin.  & -199.6822 & -199.5835 & -199.6077 & \\
   &S=1 $S_z=1$ cons.       & -199.6336 & -199.5606 & -199.5711 & \\
   & S=1 $S_z=0$ cons.      & -199.6436 & -199.5771 & -199.5776 & \\
   & S=0                    & -199.7506 & -199.6603 & -199.6553 & -199.6378 \\\hline
$\delta$ & S=1 $S_z=1$ w/o       & 0.0055 & 0.0010 & 0.0012 & \\
         & S=1 $S_z=1$ cons. pin & 0.0029 & 0.0008 & 0.0010 & \\ 
         & S=1 $S_z=1$ cons.     & 0.0038 & 0.0006 & 0.0009 & \\ 
         & S=1 $S_z=0$ cons.     & 0.0115 & 0.0006 & 0.0004 & \\
         & S=0                   & 0.0043 & 0.0010 & 0.0005 & \\\hline
$\delta_{\rm ex}$ & S=1 $S_z=1$ w/o       & 0.38 & 0.12 & 0.42 & \\ 
                  & S=1 $S_z=1$ cons. pin &	0.92 & 0.24 & 0.27 & \\ 
                  & S=1 $S_z=1$ cons.     & 0.57 & 0.08 & 0.20 & \\ 
                  & S=1 $S_z=0$ cons.     & 0.34 & 0.17 & 0.09 & \\ \hline
\end{tabular}}
\caption{\label{tab:sing-tripl} {\bf Energies of Lowest Triplet States and Singlet Ground States (in Ha) for Different RDMFT Functionals Calculated with or without Additional Constraints for $\langle \hat{\mathbf{S}}^2\rangle$.} For 2-electron systems (top), the triplet is calculated either by imposing only $S_z=1$ without additional constraints (first line for each system), or by imposing the constraint (\ref{eq:2fold}) for $S=1$, $S_z=1$ (second line for each system), or the constraint (\ref{eq:4fold}) for $S=1$, $S_z=0$ (third line). Energies for the ground state singlet are also included (fourth line). For systems with more than two electrons (bottom), the triplet is calculated with $S_z=1$ without any additional spin constraint (first line), with $S_z=1$ by imposing the constraint (\ref{eq:occM=1_pinned}) i.e., pinning the inner occupations to one (second line), and with $S_z=1$ by imposing the constraint (\ref{eq:occM=1}) (third line). For $S_z=0$, we impose the constraint (\ref{eq:occM=0_pinned}) (forth line). The exact energies obtained with MCSCF using the same basis set and the same number of active orbitals are also given for comparison. The average absolute relative deviations from MCSCF (Eq.~(\ref{eq:diff})), for the total energies, $\delta$, and the singlet-triplet excitation energies, $\delta_{\rm ex}$, are also included.}
\end{table*}

In Table \ref{tab:sing-tripl} (bottom), we include results for systems with more than two electrons. These systems are chosen because one can assume that their first excited triplet is formed by the two outermost electrons only.
 For the three functionals we considered, and in the first line for each system, we give the lowest triplet energy which is calculated using the constraint (\ref{eq:occM=1_pinned}), i.e., by pinning the occupations of all core orbitals to one and letting only the outer orbitals for the majority spin to be fractionally occupied with two electrons. This guarantees that the core electrons do not contribute to the total spin. We loosened the constraint for the inner occupations by enforcing only the constraint (\ref{eq:occM=1}). The results are given in the second line for each system. As shown in Fig.\ \ref{fig:spin_violation}, for the M\"uller functional, this leads to a deviation from the correct $\langle \hat{\mathbf{S}}^2\rangle$ but is still closer to the exact $\langle \hat{\mathbf{S}}^2\rangle$ than imposing no additional restriction.

On average the total energies of the first excited triplet states are improved imposing the constraints considered here, with the exception of the M\"uller functional with the $S=1$ and $S_z=0$ constraint. However, the singlet-to-first-triplet excitation energies worsen in many cases due to a cancellation of errors in favor of the unconstrained calculations with some functionals.

 The small number of systems that we considered, does not allow us to draw a decisive conclusion on the effect of the constraints on the excitation energies. The excitation energies (with or without the additional constraints) show a large error compared to those from MCSCF, for all the functionals we employed. This is, at least partially, due to the fact that functionals in RDMFT are typically devised and tuned to reproduce the energies of ground-state singlets. They even fail, in some cases, to identify that there is a triplet with lower energy. For example the M\"uller functional yields a singlet as the ground state for the oxygen and carbon atoms instead of the correct triplet states.

As we discussed before, with the restriction (\ref{eq:occM=1}) we cannot guarantee that we get a triplet as there exists a singlet with the same occupations. However, if the triplet is lower in energy than the corresponding singlet then the minimization will find it. The advantage of this restriction is that it can be applied for functionals that are devised to treat systems with the  same number of up and down electrons.

\section{Conclusion\label{sec:conc}}

We have considered necessary conditions for the one-body-reduced density matrix of a system of two electrons to correspond to a triplet state. There are separate conditions for the fully polarized triplet states $S_z=\pm 1$ and for the $S_z=0$ state. In a spin-restricted description, i.e., assuming the same spatial dependence of the natural orbitals in the two spin channels, the conditions for $S_z=\pm 1$ restrict the occupation numbers to be doubly degenerate. For $S_z=0$, on the other hand, a 4-fold degeneracy of the occupation numbers was found. 

We first tested if the conditions are satisfied for the fully polarized, $S_z=\pm 1$, triplet states of prototype two electron systems, namely, the helium atom and the H$_2$ molecule, using typical approximate RDMFT functionals and found that they are violated significantly. They are, however, satisfied by the exact functional for two electrons, as can be shown analytically, and in MCSCF calculations. Since the conditions only affect the degeneracy of the occupation numbers they can easily be enforced in RDMFT calculations as additional constraints in the energy minimization. Thus, we applied the conditions for $S_z=1$ to calculate the lowest excited triplet states of the aforementioned 2-electron systems. We found that, with the employed approximations, the optimal occupation numbers improve significantly compared to ``exact'' MCSCF results. We also calculated the total energies of the lowest triplet states when 
 the conditions for $S_z=1$ and $S_z=0$ are applied and we found, in most of cases, an improvement of these energies. 

We also evaluated the idea of applying the aforementioned conditions, which are exact for two 
electron systems, to systems with more than two electrons. For $S_z=\pm 1$, we employed two different approximate constraints: In the first, all electrons from the minority spin channel and all but two electrons from the majority spin channel occupy pinned natural orbitals, leaving only two electrons from the majority spin channel to lead to fractional occupation numbers. In the second, spin-up and spin-down core natural orbitals have equal occupancies which are not necessarily pinned to one, and the 2-fold degeneracy is assumed only for the weakly occupied natural orbitals which accommodate the two additional electrons of the majority spin. We evaluated the approximate constraints that we propose by testing their effect when imposed as additional constraints in RDMFT minimizations of some atoms and molecules and found that, in all cases, we get occupation numbers closer to the exact ones than without imposing them.  For $S_z=0$ triplets, the extension we considered assumes that core natural orbitals  accommodating all but two electrons form a singlet state and are pinned, while the rest of the orbitals, which accommodate the two remaining electrons follow the 4-fold degeneracy as in the case of only two electrons. Same as for the 2-electron systems, we applied the constraints both for $S_z=\pm 1$ and $S_z=0$ to RDMFT minimizations with different functionals to calculate first excited triplet states. 
 On average the constraints considered here improved the energy of the first excited triplet state, with the exception of the M\"uller functional with the $S=1$, $S_z=1$ constraints. This improvement, however, in both cases of two and more than two electrons, does not concern necessarily the corresponding singlet-triplet excitation energies. We found that for the excitation energies, error cancellations are in favor of unconstrained calculations, and the additional constraints might deteriorate the agreement of these energies with the exact. This effect is partly due to the fact that 1RDM functionals might not treat singlet and triplet states at the same level of accuracy. For the majority of present-day approximations, the main focus has been the accurate description of singlet states and there is no guarantee of the quality of their results when extended to triplet states. For example, most functionals break the degeneracy between the fully polarized and the $S_z=0$ triplet states.

The present work is a significant step in the description of high-spin states using reduced density matrix functional theory. Our findings motivate the development of approximations which could offer a better description for triplet states by following the proposed recipe. A benchmark for these approximations would be the prediction of the triplet ground states of atomic and molecular systems. Finally, with the proposed methodology, it becomes feasible to access the $S_z=0$ triplet state in RDMFT by applying the appropriate necessary conditions. Consequently, for any new functional one could test the degeneracy between the fully polarized and the $S_z=0$ triplet states. In the future, with the improvement of available approximations, it will be possible for RDMFT to study cases of broken degeneracy of the triplet states, e.g.\ when magnetic fields are applied.

\section{Acknowledgments}
IT and NH acknowledge support from a Emmy-Noether grant from Deutsche Forschungsgemeinschaft. NNL acknowledges support from the Greek Ministry of Education (E$\Sigma\Pi$A program), GSRT action $\rm KPH\Pi I\Sigma$, project ``New multifunctional Nanostructured Materials and Devices - POLYNANO'' (No. 447963). IT would like to acknowledge Dr. S.Thanos for useful discussions on the manuscript. 

\begin{appendix}
\section{Degeneracies in the M\"uller Functional}\label{app:mueller}
In this Appendix, we show that for 2-electron systems, the M\"uller functional
respects the energy degeneracy of the ${S}_z=1$ and ${S}_z=0$ triplet states. The states (\ref{eq:trips=1}) and (\ref{eq:trips=0}) have the same natural orbitals but differ in their occupation. If the occupation numbers of the ${S}_z=1$ state are denoted by $n_{j\uparrow}$ (the down channel is empty) then the occupation numbers for the ${S}_z=0$ state, $\tilde{n}_{j\sigma}$, are given by
\be\label{eq:occtrips=0}
\tilde{n}_{j\uparrow}=\tilde{n}_{j\downarrow}=n_{j\uparrow}/2.
\ee
Starting from the solution of the fully polarized state, we know that the total energy is given by 
\bea
&&E = \sum_{j=1}^\infty n_{j\uparrow}\int d^3r \varphi_j^*(\v r)\left(-\frac{\nabla^2}{2}+v_\mathrm{ext}(\v r)\right)\varphi_j(\v r) \nonumber\\
&&+\frac{1}{2}\sum_{j,k=1}^\infty n_{j\uparrow}n_{k\uparrow} J_{jk}-\frac{1}{2}\sum_{j,k=1}^\infty \sqrt{n_{j\uparrow}n_{k\uparrow}} K_{jk}
\label{eq:energytrips=1}
\eea
with
\bea
J_{jk}&=&\iint d^3r d^3r'\frac{|\varphi_j(\v r)|^2|\varphi_k(\v r')|^2}{|\v r-\v r'|},\\
K_{jk}&=&\iint d^3r d^3r'\frac{\varphi_j^*(\v r)\varphi_k^*(\v r')\varphi_k(\v r)\varphi_j(\v r')}{|\v r-\v r'|}.
\eea
The derivative of the total energy with respect to the occupation number $n_{j\uparrow}$ reads as
\bea
\frac{\partial E}{\partial n_{j\uparrow}}&=&\int d^3r \varphi_j^*(\v r)\left(-\frac{\nabla^2}{2}+v_\mathrm{ext}(\v r)\right)\varphi_j(\v r)\nonumber \\
&+&\sum_{k=1}^\infty n_{k\uparrow} J_{jk}-\sum_{k=1}^\infty \frac{\sqrt{n_{k\uparrow}}}{2\sqrt{n_{j\uparrow}}} K_{jk}.
\label{eq:derivs=1}
\eea
As we are at the solution point, the derivatives with respect to all fractional occupation numbers satisfy 
\be
\frac{\partial E}{\partial n_{j\uparrow}}=\mu,
\ee
where $\mu$ denotes the chemical potential of the system.

Using the occupation numbers $\tilde{n}_{j\sigma}$ of the $\langle\hat{S}_z\rangle=0$ state instead, the total energy is given as
\bea
E &=& \sum_{\sigma}\sum_{j=1}^\infty \tilde{n}_{j\sigma}\int d^3r \varphi_j^*(\v r)\left(-\frac{\nabla^2}{2}+v_\mathrm{ext}(\v r)\right)\varphi_j(\v r) \nonumber\\
&+&\frac{1}{2}\sum_{\sigma\sigma'}\sum_{j,k=1}^\infty \tilde{n}_{j\sigma}\tilde{n}_{k\sigma'} J_{jk}\nonumber\\
&-&\frac{1}{2}\sum_\sigma\sum_{j,k=1}^\infty \sqrt{\tilde{n}_{j\sigma}\tilde{n}_{k\sigma}} K_{jk}.
\eea
Making the spin sums explicit this can be rewritten as
\bea
E &=& \sum_{j=1}^\infty \left(\tilde{n}_{j\uparrow}+\tilde{n}_{j\downarrow}\right)
\int d^3r \varphi_j^*(\v r)\left(-\frac{\nabla^2}{2}+v_\mathrm{ext}(\v r)\right)\varphi_j(\v r) \nonumber\\
&+&\frac{1}{2}\sum_{j,k=1}^\infty 
\left(\tilde{n}_{j\uparrow}+\tilde{n}_{j\downarrow}\right)\left(\tilde{n}_{k\uparrow}+\tilde{n}_{k\downarrow}\right) J_{jk} \nonumber\\
&-&\frac{1}{2}\sum_{j,k=1}^\infty 
\left(\sqrt{\tilde{n}_{j\uparrow}\tilde{n}_{k\uparrow}}+\sqrt{\tilde{n}_{j\downarrow}\tilde{n}_{k\downarrow}}\right) K_{jk}
\eea
which, using Eq.\ (\ref{eq:occtrips=0}) is identical to the energy of the $\langle\hat{S}_z\rangle=1$ state, Eq.\ (\ref{eq:energytrips=1}).
However, we still need to show that this energy is also an extremum. For the derivative with respect to the occupation numbers we obtain
\bea
\frac{\partial E}{\partial \tilde{n}_{j\sigma}}&=&\int d^3r \varphi_j^*(\v r)\left(-\frac{\nabla^2}{2}+v_\mathrm{ext}(\v r)\right)\varphi_j(\v r)\nonumber \\
&+&\sum_{\sigma'}\sum_{k=1}^\infty \tilde{n}_{k\sigma'} J_{jk}-\sum_{k=1}^\infty \frac{\sqrt{\tilde{n}_{k\sigma}}}{2\sqrt{\tilde{n}_{j\sigma}}} K_{jk}.
\eea
Again making the spin sum explicit in the second term and using $\tilde{n}_{k\sigma}/\tilde{n}_{j\sigma}=n_{k\uparrow}/n_{j\uparrow}$ we find that the derivative is the same as in Eq.\ (\ref{eq:derivs=1}). Hence, if the occupation numbers $n_{j\uparrow}$ minimize the total energy for the triplet ${S}_z=1$ state then the occupation numbers $\tilde{n}_{j\sigma}$ defined in Eq.\ (\ref{eq:occtrips=0}) form the minimum for the triplet ${S}_z=0$ state.

We note that this derivation crucially depends on the square root dependence in the exchange term. For a general power $\alpha$, i.e., $(n_{j\sigma}n_{k\sigma})^\alpha$ in the exchange energy, one finds factors of $2/2^{2\alpha}$ and $1/2^{2\alpha-1}$ in the exchange energy and its derivative, respectively, when comparing the terms for $n_{j\uparrow}$ and $\tilde{n}_{j\sigma}$. In other words, the terms are only the same for $\alpha=1/2$. 

We also emphasize that the degeneracy holds only if the two sets of orbitals are identical. Since the two states (\ref{eq:trips=1}) and (\ref{eq:trips=0}) are connected by $\hat{S}_\pm$, which only acts on the spin degrees of freedom, this is satisfied. However, if one determines the orbitals from an energy minimization it can happen that one finds different minima for the orbitals in the two cases resulting in a broken degeneracy. In the systems that we have tested, we found that the degeneracy is satisfied with an accuracy of 6 decimal digits which corresponds to the convergence of the overall calculation. 

As we have seen, for the M\"uller functional the sums are the same because all the terms in the sums are identical. For the other functionals considered here, there might be specific cases of sets of occupation numbers that triplet states are degenerate, i.e., the sums are equal, however this degeneracy does not hold in general.

\end{appendix}

\bibliographystyle{apsrev}

\begin{thebibliography}{51}
\expandafter\ifx\csname natexlab\endcsname\relax\def\natexlab#1{#1}\fi
\expandafter\ifx\csname bibnamefont\endcsname\relax
  \def\bibnamefont#1{#1}\fi
\expandafter\ifx\csname bibfnamefont\endcsname\relax
  \def\bibfnamefont#1{#1}\fi
\expandafter\ifx\csname citenamefont\endcsname\relax
  \def\citenamefont#1{#1}\fi
\expandafter\ifx\csname url\endcsname\relax
  \def\url#1{\texttt{#1}}\fi
\expandafter\ifx\csname urlprefix\endcsname\relax\def\urlprefix{URL }\fi
\providecommand{\bibinfo}[2]{#2}
\providecommand{\eprint}[2][]{\url{#2}}

\bibitem[{\citenamefont{Gilbert}(1975)}]{G1975}
\bibinfo{author}{\bibfnamefont{T.~L.} \bibnamefont{Gilbert}},
  \bibinfo{journal}{Phys. Rev. B} \textbf{\bibinfo{volume}{12}},
  \bibinfo{pages}{2111} (\bibinfo{year}{1975}).

\bibitem[{\citenamefont{Lathiotakis et~al.}(2005)\citenamefont{Lathiotakis,
  Helbig, and Gross}}]{LHG2005}
\bibinfo{author}{\bibfnamefont{N.~N.} \bibnamefont{Lathiotakis}},
  \bibinfo{author}{\bibfnamefont{N.}~\bibnamefont{Helbig}}, \bibnamefont{and}
  \bibinfo{author}{\bibfnamefont{E.~K.~U.} \bibnamefont{Gross}},
  \bibinfo{journal}{Phys. Rev. A} \textbf{\bibinfo{volume}{72}},
  \bibinfo{pages}{030501} (\bibinfo{year}{2005}).

\bibitem[{\citenamefont{Piris et~al.}(2009)\citenamefont{Piris, Matxain, Lopez,
  and Ugalde}}]{Piris1}
\bibinfo{author}{\bibfnamefont{M.}~\bibnamefont{Piris}},
  \bibinfo{author}{\bibfnamefont{J.~M.} \bibnamefont{Matxain}},
  \bibinfo{author}{\bibfnamefont{X.}~\bibnamefont{Lopez}}, \bibnamefont{and}
  \bibinfo{author}{\bibfnamefont{J.~M.} \bibnamefont{Ugalde}},
  \bibinfo{journal}{The Journal of Chemical Physics}
  \textbf{\bibinfo{volume}{131}}, \bibinfo{pages}{021102}
  (\bibinfo{year}{2009}).

\bibitem[{\citenamefont{Rohr and Pernal}(2011)}]{RP2011}
\bibinfo{author}{\bibfnamefont{D.~R.} \bibnamefont{Rohr}} \bibnamefont{and}
  \bibinfo{author}{\bibfnamefont{K.}~\bibnamefont{Pernal}},
  \bibinfo{journal}{The Journal of Chemical Physics}
  \textbf{\bibinfo{volume}{135}}, \bibinfo{pages}{074104}
  (\bibinfo{year}{2011}).

\bibitem[{\citenamefont{Leiva and Piris}(2007)}]{Piris_Leiva_spin}
\bibinfo{author}{\bibfnamefont{P.}~\bibnamefont{Leiva}} \bibnamefont{and}
  \bibinfo{author}{\bibfnamefont{M.}~\bibnamefont{Piris}},
  \bibinfo{journal}{International Journal of Quantum Chemistry}
  \textbf{\bibinfo{volume}{107}}, \bibinfo{pages}{1} (\bibinfo{year}{2007}).

\bibitem[{\citenamefont{Hohenberg and Kohn}(1964)}]{HK1964}
\bibinfo{author}{\bibfnamefont{P.}~\bibnamefont{Hohenberg}} \bibnamefont{and}
  \bibinfo{author}{\bibfnamefont{W.}~\bibnamefont{Kohn}},
  \bibinfo{journal}{Phys. Rev.} \textbf{\bibinfo{volume}{136}},
  \bibinfo{pages}{B864} (\bibinfo{year}{1964}).

\bibitem[{\citenamefont{Pernal and Giesbertz}(2016)}]{Pernal2016}
\bibinfo{author}{\bibfnamefont{K.}~\bibnamefont{Pernal}} \bibnamefont{and}
  \bibinfo{author}{\bibfnamefont{K.~J.~H.} \bibnamefont{Giesbertz}}, in
  \emph{\bibinfo{booktitle}{Density-Functional Methods for Excited States}},
  edited by \bibinfo{editor}{\bibfnamefont{N.}~\bibnamefont{Ferr{\'e}}},
  \bibinfo{editor}{\bibfnamefont{M.}~\bibnamefont{Filatov}}, \bibnamefont{and}
  \bibinfo{editor}{\bibfnamefont{M.}~\bibnamefont{Huix-Rotllant}}
  (\bibinfo{publisher}{Springer International Publishing},
  \bibinfo{address}{Cham}, \bibinfo{year}{2016}), pp.
  \bibinfo{pages}{125--183}, ISBN \bibinfo{isbn}{978-3-319-22081-9}.

\bibitem[{\citenamefont{M\"uller}(1984)}]{M1984}
\bibinfo{author}{\bibfnamefont{A.~M.~K.} \bibnamefont{M\"uller}},
  \bibinfo{journal}{Phys. Lett. A} \textbf{\bibinfo{volume}{105}},
  \bibinfo{pages}{446} (\bibinfo{year}{1984}).

\bibitem[{\citenamefont{Goedecker and Umrigar}(1998)}]{GU1998}
\bibinfo{author}{\bibfnamefont{S.}~\bibnamefont{Goedecker}} \bibnamefont{and}
  \bibinfo{author}{\bibfnamefont{C.~J.} \bibnamefont{Umrigar}},
  \bibinfo{journal}{Phys. Rev. Lett.} \textbf{\bibinfo{volume}{81}},
  \bibinfo{pages}{866} (\bibinfo{year}{1998}).

\bibitem[{\citenamefont{Buijse and Baerends}(2002)}]{BB2002}
\bibinfo{author}{\bibfnamefont{M.~A.} \bibnamefont{Buijse}} \bibnamefont{and}
  \bibinfo{author}{\bibfnamefont{E.~J.} \bibnamefont{Baerends}},
  \bibinfo{journal}{Mol. Phys.} \textbf{\bibinfo{volume}{100}},
  \bibinfo{pages}{401} (\bibinfo{year}{2002}).

\bibitem[{\citenamefont{Gritsenko et~al.}(2005)\citenamefont{Gritsenko, Pernal,
  and Baerends}}]{GPB2005}
\bibinfo{author}{\bibfnamefont{O.}~\bibnamefont{Gritsenko}},
  \bibinfo{author}{\bibfnamefont{K.}~\bibnamefont{Pernal}}, \bibnamefont{and}
  \bibinfo{author}{\bibfnamefont{E.~J.} \bibnamefont{Baerends}},
  \bibinfo{journal}{J. Chem. Phys.} \textbf{\bibinfo{volume}{122}},
  \bibinfo{pages}{204102} (\bibinfo{year}{2005}).

\bibitem[{\citenamefont{Lathiotakis
  et~al.}(2009{\natexlab{a}})\citenamefont{Lathiotakis, Helbig, Zacarias, and
  Gross}}]{LHZG2009}
\bibinfo{author}{\bibfnamefont{N.~N.} \bibnamefont{Lathiotakis}},
  \bibinfo{author}{\bibfnamefont{N.}~\bibnamefont{Helbig}},
  \bibinfo{author}{\bibfnamefont{A.}~\bibnamefont{Zacarias}}, \bibnamefont{and}
  \bibinfo{author}{\bibfnamefont{E.~K.~U.} \bibnamefont{Gross}},
  \bibinfo{journal}{The Journal of Chemical Physics}
  \textbf{\bibinfo{volume}{130}}, \bibinfo{pages}{064109}
  (\bibinfo{year}{2009}{\natexlab{a}}).

\bibitem[{\citenamefont{Piris}(2014)}]{P2014}
\bibinfo{author}{\bibfnamefont{M.}~\bibnamefont{Piris}}, \bibinfo{journal}{J.
  Chem. Phys.} \textbf{\bibinfo{volume}{141}}, \bibinfo{pages}{044107}
  (\bibinfo{year}{2014}).

\bibitem[{\citenamefont{Pernal}(2014)}]{KP2014}
\bibinfo{author}{\bibfnamefont{K.}~\bibnamefont{Pernal}}, \bibinfo{journal}{J.
  Chem. Theory Comput.} \textbf{\bibinfo{volume}{10}}, \bibinfo{pages}{4332}
  (\bibinfo{year}{2014}).

\bibitem[{\citenamefont{Rohr et~al.}(2008)\citenamefont{Rohr, Pernal,
  Gritsenko, and Baerends}}]{RPGB2008}
\bibinfo{author}{\bibfnamefont{D.~R.} \bibnamefont{Rohr}},
  \bibinfo{author}{\bibfnamefont{K.}~\bibnamefont{Pernal}},
  \bibinfo{author}{\bibfnamefont{O.~V.} \bibnamefont{Gritsenko}},
  \bibnamefont{and} \bibinfo{author}{\bibfnamefont{E.~J.}
  \bibnamefont{Baerends}}, \bibinfo{journal}{J. Chem. Phys.}
  \textbf{\bibinfo{volume}{129}}, \bibinfo{pages}{164105}
  (\bibinfo{year}{2008}).

\bibitem[{\citenamefont{Marques and Lathiotakis}(2008)}]{ML2008}
\bibinfo{author}{\bibfnamefont{M.~A.~L.} \bibnamefont{Marques}}
  \bibnamefont{and} \bibinfo{author}{\bibfnamefont{N.~N.}
  \bibnamefont{Lathiotakis}}, \bibinfo{journal}{Phys. Rev. A}
  \textbf{\bibinfo{volume}{77}}, \bibinfo{pages}{032509}
  (\bibinfo{year}{2008}).

\bibitem[{\citenamefont{Sharma et~al.}(2008)\citenamefont{Sharma, Dewhurst,
  Lathiotakis, and Gross}}]{SDLG2008}
\bibinfo{author}{\bibfnamefont{S.}~\bibnamefont{Sharma}},
  \bibinfo{author}{\bibfnamefont{J.~K.} \bibnamefont{Dewhurst}},
  \bibinfo{author}{\bibfnamefont{N.~N.} \bibnamefont{Lathiotakis}},
  \bibnamefont{and} \bibinfo{author}{\bibfnamefont{E.~K.~U.}
  \bibnamefont{Gross}}, \bibinfo{journal}{Phys. Rev. B}
  \textbf{\bibinfo{volume}{78}}, \bibinfo{pages}{201103}
  (\bibinfo{year}{2008}).

\bibitem[{\citenamefont{Shinohara et~al.}(2015)\citenamefont{Shinohara, Sharma,
  Shallcross, Lathiotakis, and Gross}}]{SSSLG2015}
\bibinfo{author}{\bibfnamefont{Y.}~\bibnamefont{Shinohara}},
  \bibinfo{author}{\bibfnamefont{S.}~\bibnamefont{Sharma}},
  \bibinfo{author}{\bibfnamefont{S.}~\bibnamefont{Shallcross}},
  \bibinfo{author}{\bibfnamefont{N.~N.} \bibnamefont{Lathiotakis}},
  \bibnamefont{and} \bibinfo{author}{\bibfnamefont{E.~K.~U.}
  \bibnamefont{Gross}}, \bibinfo{journal}{Journal of Chemical Theory and
  Computation} \textbf{\bibinfo{volume}{11}}, \bibinfo{pages}{4895}
  (\bibinfo{year}{2015}).

\bibitem[{\citenamefont{Lathiotakis
  et~al.}(2009{\natexlab{b}})\citenamefont{Lathiotakis, Sharma, Dewhurst, Eich,
  Marques, and Gross}}]{LSDEMG2009}
\bibinfo{author}{\bibfnamefont{N.}~\bibnamefont{Lathiotakis}},
  \bibinfo{author}{\bibfnamefont{S.}~\bibnamefont{Sharma}},
  \bibinfo{author}{\bibfnamefont{J.}~\bibnamefont{Dewhurst}},
  \bibinfo{author}{\bibfnamefont{F.}~\bibnamefont{Eich}},
  \bibinfo{author}{\bibfnamefont{M.}~\bibnamefont{Marques}}, \bibnamefont{and}
  \bibinfo{author}{\bibfnamefont{E.}~\bibnamefont{Gross}},
  \bibinfo{journal}{Phys. Rev. A} \textbf{\bibinfo{volume}{79}},
  \bibinfo{pages}{040501} (\bibinfo{year}{2009}{\natexlab{b}}).

\bibitem[{\citenamefont{Piris}(2006)}]{P2006}
\bibinfo{author}{\bibfnamefont{M.}~\bibnamefont{Piris}}, \bibinfo{journal}{Int.
  J. Quant. Chem.} \textbf{\bibinfo{volume}{106}}, \bibinfo{pages}{1093}
  (\bibinfo{year}{2006}).

\bibitem[{\citenamefont{Piris et~al.}(2010)\citenamefont{Piris, Matxain, Lopez,
  and Ugalde}}]{PMLU2010}
\bibinfo{author}{\bibfnamefont{M.}~\bibnamefont{Piris}},
  \bibinfo{author}{\bibfnamefont{J.~M.} \bibnamefont{Matxain}},
  \bibinfo{author}{\bibfnamefont{X.}~\bibnamefont{Lopez}}, \bibnamefont{and}
  \bibinfo{author}{\bibfnamefont{J.~M.} \bibnamefont{Ugalde}},
  \bibinfo{journal}{J. Chem. Phys.} \textbf{\bibinfo{volume}{132}},
  \bibinfo{pages}{031103} (\bibinfo{year}{2010}).

\bibitem[{\citenamefont{Piris et~al.}(2011)\citenamefont{Piris, Lopez,
  Ruiperez, Matxain, and Ugalde}}]{PLRMU2011}
\bibinfo{author}{\bibfnamefont{M.}~\bibnamefont{Piris}},
  \bibinfo{author}{\bibfnamefont{X.}~\bibnamefont{Lopez}},
  \bibinfo{author}{\bibfnamefont{F.}~\bibnamefont{Ruiperez}},
  \bibinfo{author}{\bibfnamefont{J.~M.} \bibnamefont{Matxain}},
  \bibnamefont{and} \bibinfo{author}{\bibfnamefont{J.~M.}
  \bibnamefont{Ugalde}}, \bibinfo{journal}{J. Chem. Phys.}
  \textbf{\bibinfo{volume}{134}}, \bibinfo{pages}{164102}
  (\bibinfo{year}{2011}).

\bibitem[{\citenamefont{Surj\'an}(1999)}]{S1999}
\bibinfo{author}{\bibfnamefont{P.~R.} \bibnamefont{Surj\'an}}, in
  \emph{\bibinfo{booktitle}{Correlation and Localization}}, edited by
  \bibinfo{editor}{\bibfnamefont{P.~R.} \bibnamefont{Surj\'an}},
  \bibinfo{editor}{\bibfnamefont{R.~J.} \bibnamefont{Bartlett}},
  \bibinfo{editor}{\bibfnamefont{F.}~\bibnamefont{Bog\'ar}},
  \bibinfo{editor}{\bibfnamefont{D.~L.} \bibnamefont{Cooper}},
  \bibinfo{editor}{\bibfnamefont{B.}~\bibnamefont{Kirtman}},
  \bibinfo{editor}{\bibfnamefont{W.}~\bibnamefont{Klopper}},
  \bibinfo{editor}{\bibfnamefont{W.}~\bibnamefont{Kutzelnigg}},
  \bibinfo{editor}{\bibfnamefont{N.~H.} \bibnamefont{March}},
  \bibinfo{editor}{\bibfnamefont{P.~G.} \bibnamefont{Mezey}},
  \bibinfo{editor}{\bibfnamefont{H.}~\bibnamefont{M\"uller}},
  \bibnamefont{et~al.} (\bibinfo{publisher}{Springer Berlin Heidelberg},
  \bibinfo{year}{1999}), vol. \bibinfo{volume}{203} of
  \emph{\bibinfo{series}{Topics in Current Chemistry}}, pp.
  \bibinfo{pages}{63--88}, ISBN \bibinfo{isbn}{978-3-540-65754-5}.

\bibitem[{\citenamefont{Rassolov}(2002)}]{R2002}
\bibinfo{author}{\bibfnamefont{V.}~\bibnamefont{Rassolov}},
  \bibinfo{journal}{J. Chem. Phys.} \textbf{\bibinfo{volume}{117}},
  \bibinfo{pages}{5978} (\bibinfo{year}{2002}).

\bibitem[{\citenamefont{Rassolov and Xu}(2007)}]{RX2007}
\bibinfo{author}{\bibfnamefont{V.~A.} \bibnamefont{Rassolov}} \bibnamefont{and}
  \bibinfo{author}{\bibfnamefont{F.}~\bibnamefont{Xu}}, \bibinfo{journal}{J.
  Chem. Phys.} \textbf{\bibinfo{volume}{127}}, \bibinfo{pages}{044104}
  (\bibinfo{year}{2007}).

\bibitem[{\citenamefont{Lathiotakis et~al.}(2014)\citenamefont{Lathiotakis,
  Helbig, Rubio, and Gidopoulos}}]{LHRG2014}
\bibinfo{author}{\bibfnamefont{N.~N.} \bibnamefont{Lathiotakis}},
  \bibinfo{author}{\bibfnamefont{N.}~\bibnamefont{Helbig}},
  \bibinfo{author}{\bibfnamefont{A.}~\bibnamefont{Rubio}}, \bibnamefont{and}
  \bibinfo{author}{\bibfnamefont{N.~I.} \bibnamefont{Gidopoulos}},
  \bibinfo{journal}{Phys. Rev. A} \textbf{\bibinfo{volume}{90}},
  \bibinfo{pages}{032511} (\bibinfo{year}{2014}).

\bibitem[{\citenamefont{Matxain et~al.}(2011)\citenamefont{Matxain, Piris,
  Ruiperez, Lopez, and Ugalde}}]{MPRLU2011}
\bibinfo{author}{\bibfnamefont{J.~M.} \bibnamefont{Matxain}},
  \bibinfo{author}{\bibfnamefont{M.}~\bibnamefont{Piris}},
  \bibinfo{author}{\bibfnamefont{F.}~\bibnamefont{Ruiperez}},
  \bibinfo{author}{\bibfnamefont{X.}~\bibnamefont{Lopez}}, \bibnamefont{and}
  \bibinfo{author}{\bibfnamefont{J.~M.} \bibnamefont{Ugalde}},
  \bibinfo{journal}{Phys. Chem. Chem. Phys.} \textbf{\bibinfo{volume}{13}},
  \bibinfo{pages}{20129} (\bibinfo{year}{2011}).

\bibitem[{\citenamefont{Helbig et~al.}(2007)\citenamefont{Helbig, Lathiotakis,
  Albrecht, and Gross}}]{HLAG2007}
\bibinfo{author}{\bibfnamefont{N.}~\bibnamefont{Helbig}},
  \bibinfo{author}{\bibfnamefont{N.~N.} \bibnamefont{Lathiotakis}},
  \bibinfo{author}{\bibfnamefont{M.}~\bibnamefont{Albrecht}}, \bibnamefont{and}
  \bibinfo{author}{\bibfnamefont{E.~K.~U.} \bibnamefont{Gross}},
  \bibinfo{journal}{EPL (Europhysics Letters)} \textbf{\bibinfo{volume}{77}},
  \bibinfo{pages}{67003} (\bibinfo{year}{2007}).

\bibitem[{\citenamefont{Lathiotakis et~al.}(2010)\citenamefont{Lathiotakis,
  Sharma, Helbig, Dewhurst, Marques, Baldsiefen, Zacarias, and
  Gross}}]{Letc2010}
\bibinfo{author}{\bibfnamefont{N.~N.} \bibnamefont{Lathiotakis}},
  \bibinfo{author}{\bibfnamefont{S.}~\bibnamefont{Sharma}},
  \bibinfo{author}{\bibfnamefont{N.}~\bibnamefont{Helbig}},
  \bibinfo{author}{\bibfnamefont{J.}~\bibnamefont{Dewhurst}},
  \bibinfo{author}{\bibfnamefont{M.}~\bibnamefont{Marques}},
  \bibinfo{author}{\bibfnamefont{T.}~\bibnamefont{Baldsiefen}},
  \bibinfo{author}{\bibfnamefont{A.}~\bibnamefont{Zacarias}}, \bibnamefont{and}
  \bibinfo{author}{\bibfnamefont{E.}~\bibnamefont{Gross}},
  \bibinfo{journal}{Zeitscrift f\"ur Physikalische Chemie}
  \textbf{\bibinfo{volume}{224}}, \bibinfo{pages}{467} (\bibinfo{year}{2010}).

\bibitem[{\citenamefont{Putaja et~al.}(2016)\citenamefont{Putaja, Eich,
  Baldsiefen, and R\"as\"anen}}]{Constraints_Power}
\bibinfo{author}{\bibfnamefont{A.}~\bibnamefont{Putaja}},
  \bibinfo{author}{\bibfnamefont{F.~G.} \bibnamefont{Eich}},
  \bibinfo{author}{\bibfnamefont{T.}~\bibnamefont{Baldsiefen}},
  \bibnamefont{and}
  \bibinfo{author}{\bibfnamefont{E.}~\bibnamefont{R\"as\"anen}},
  \bibinfo{journal}{Phys. Rev. A} \textbf{\bibinfo{volume}{93}},
  \bibinfo{pages}{032503} (\bibinfo{year}{2016}),
  \urlprefix\url{http://link.aps.org/doi/10.1103/PhysRevA.93.032503}.

\bibitem[{\citenamefont{Coleman}(1963)}]{Coleman_1963}
\bibinfo{author}{\bibfnamefont{A.~J.} \bibnamefont{Coleman}},
  \bibinfo{journal}{Rev. Mod. Phys.} \textbf{\bibinfo{volume}{35}},
  \bibinfo{pages}{668} (\bibinfo{year}{1963}).

\bibitem[{\citenamefont{Smith}(1966)}]{S1966}
\bibinfo{author}{\bibfnamefont{D.~W.} \bibnamefont{Smith}},
  \bibinfo{journal}{Phys. Rev.} \textbf{\bibinfo{volume}{147}},
  \bibinfo{pages}{896} (\bibinfo{year}{1966}).

\bibitem[{\citenamefont{Altunbulak}(2008)}]{ThesisAltunbulak}
\bibinfo{author}{\bibfnamefont{M.}~\bibnamefont{Altunbulak}}, Ph.D. thesis,
  \bibinfo{school}{Bilkent University} (\bibinfo{year}{2008}).

\bibitem[{\citenamefont{Altunbulak and Klyachko}(2008)}]{Klyachko_Math}
\bibinfo{author}{\bibfnamefont{M.}~\bibnamefont{Altunbulak}} \bibnamefont{and}
  \bibinfo{author}{\bibfnamefont{A.}~\bibnamefont{Klyachko}},
  \bibinfo{journal}{Communications in Mathematical Physics}
  \textbf{\bibinfo{volume}{282}}, \bibinfo{pages}{287} (\bibinfo{year}{2008}).

\bibitem[{\citenamefont{Borland and Dennis}(1972)}]{Borland-Dennis}
\bibinfo{author}{\bibfnamefont{R.~E.} \bibnamefont{Borland}} \bibnamefont{and}
  \bibinfo{author}{\bibfnamefont{K.}~\bibnamefont{Dennis}},
  \bibinfo{journal}{Journal of Physics B: Atomic and Molecular Physics}
  \textbf{\bibinfo{volume}{5}}, \bibinfo{pages}{7} (\bibinfo{year}{1972}).

\bibitem[{\citenamefont{Schilling et~al.}(2013)\citenamefont{Schilling, Gross,
  and Christandl}}]{PRL_pure_cond}
\bibinfo{author}{\bibfnamefont{C.}~\bibnamefont{Schilling}},
  \bibinfo{author}{\bibfnamefont{D.}~\bibnamefont{Gross}}, \bibnamefont{and}
  \bibinfo{author}{\bibfnamefont{M.}~\bibnamefont{Christandl}},
  \bibinfo{journal}{Phys. Rev. Lett.} \textbf{\bibinfo{volume}{110}},
  \bibinfo{pages}{040404} (\bibinfo{year}{2013}).

\bibitem[{\citenamefont{Benavides-Riveros
  et~al.}(2013)\citenamefont{Benavides-Riveros, Gracia-Bond\'{\i}a, and
  Springborg}}]{Carlos1}
\bibinfo{author}{\bibfnamefont{C.~L.} \bibnamefont{Benavides-Riveros}},
  \bibinfo{author}{\bibfnamefont{J.~M.} \bibnamefont{Gracia-Bond\'{\i}a}},
  \bibnamefont{and}
  \bibinfo{author}{\bibfnamefont{M.}~\bibnamefont{Springborg}},
  \bibinfo{journal}{Phys. Rev. A} \textbf{\bibinfo{volume}{88}},
  \bibinfo{pages}{022508} (\bibinfo{year}{2013}).

\bibitem[{\citenamefont{Benavides-Riveros and Springborg}(2015)}]{Carlos2}
\bibinfo{author}{\bibfnamefont{C.~L.} \bibnamefont{Benavides-Riveros}}
  \bibnamefont{and}
  \bibinfo{author}{\bibfnamefont{M.}~\bibnamefont{Springborg}},
  \bibinfo{journal}{Phys. Rev. A} \textbf{\bibinfo{volume}{92}},
  \bibinfo{pages}{012512} (\bibinfo{year}{2015}).

\bibitem[{\citenamefont{Schilling}(2015)}]{Schilling_2015}
\bibinfo{author}{\bibfnamefont{C.}~\bibnamefont{Schilling}},
  \bibinfo{journal}{Phys. Rev. A} \textbf{\bibinfo{volume}{91}},
  \bibinfo{pages}{022105} (\bibinfo{year}{2015}).

\bibitem[{\citenamefont{Theophilou et~al.}(2015)\citenamefont{Theophilou,
  Lathiotakis, Marques, and Helbig}}]{TLMH2015}
\bibinfo{author}{\bibfnamefont{I.}~\bibnamefont{Theophilou}},
  \bibinfo{author}{\bibfnamefont{N.~N.} \bibnamefont{Lathiotakis}},
  \bibinfo{author}{\bibfnamefont{M.~A.~L.} \bibnamefont{Marques}},
  \bibnamefont{and} \bibinfo{author}{\bibfnamefont{N.}~\bibnamefont{Helbig}},
  \bibinfo{journal}{The Journal of Chemical Physics}
  \textbf{\bibinfo{volume}{142}}, \bibinfo{pages}{154108}
  (\bibinfo{year}{2015}).

\bibitem[{\citenamefont{L\"owdin}(1955)}]{L1955}
\bibinfo{author}{\bibfnamefont{P.-O.} \bibnamefont{L\"owdin}},
  \bibinfo{journal}{Phys. Rev.} \textbf{\bibinfo{volume}{97}},
  \bibinfo{pages}{1474} (\bibinfo{year}{1955}).

\bibitem[{\citenamefont{Alcoba and Valdemoro}(2005)}]{AV2005}
\bibinfo{author}{\bibfnamefont{D.~R.} \bibnamefont{Alcoba}} \bibnamefont{and}
  \bibinfo{author}{\bibfnamefont{C.}~\bibnamefont{Valdemoro}},
  \bibinfo{journal}{International Journal of Quantum Chemistry}
  \textbf{\bibinfo{volume}{102}}, \bibinfo{pages}{629} (\bibinfo{year}{2005}),
  ISSN \bibinfo{issn}{1097-461X}.

\bibitem[{\citenamefont{Ramos-Cordoba et~al.}(2014)\citenamefont{Ramos-Cordoba,
  Salvador, Piris, and Matito}}]{Piris2}
\bibinfo{author}{\bibfnamefont{E.}~\bibnamefont{Ramos-Cordoba}},
  \bibinfo{author}{\bibfnamefont{P.}~\bibnamefont{Salvador}},
  \bibinfo{author}{\bibfnamefont{M.}~\bibnamefont{Piris}}, \bibnamefont{and}
  \bibinfo{author}{\bibfnamefont{E.}~\bibnamefont{Matito}},
  \bibinfo{journal}{The Journal of Chemical Physics}
  \textbf{\bibinfo{volume}{141}}, \bibinfo{pages}{234101}
  (\bibinfo{year}{2014}).

\bibitem[{\citenamefont{Cohen et~al.}(2007)\citenamefont{Cohen, Tozer, and
  Handy}}]{Handy_spin_dft}
\bibinfo{author}{\bibfnamefont{A.~J.} \bibnamefont{Cohen}},
  \bibinfo{author}{\bibfnamefont{D.~J.} \bibnamefont{Tozer}}, \bibnamefont{and}
  \bibinfo{author}{\bibfnamefont{N.~C.} \bibnamefont{Handy}},
  \bibinfo{journal}{The Journal of Chemical Physics}
  \textbf{\bibinfo{volume}{126}}, \bibinfo{pages}{214104}
  (\bibinfo{year}{2007}).

\bibitem[{\citenamefont{Jacob and Reiher}(2012)}]{spin_dft_reiher}
\bibinfo{author}{\bibfnamefont{C.~R.} \bibnamefont{Jacob}} \bibnamefont{and}
  \bibinfo{author}{\bibfnamefont{M.}~\bibnamefont{Reiher}},
  \bibinfo{journal}{International Journal of Quantum Chemistry}
  \textbf{\bibinfo{volume}{112}}, \bibinfo{pages}{3661} (\bibinfo{year}{2012}),
  ISSN \bibinfo{issn}{1097-461X}.

\bibitem[{\citenamefont{Wang et~al.}(1995)\citenamefont{Wang, Becke, and
  Smith}}]{spin_becke}
\bibinfo{author}{\bibfnamefont{J.}~\bibnamefont{Wang}},
  \bibinfo{author}{\bibfnamefont{A.~D.} \bibnamefont{Becke}}, \bibnamefont{and}
  \bibinfo{author}{\bibfnamefont{V.~H.} \bibnamefont{Smith}},
  \bibinfo{journal}{The Journal of Chemical Physics}
  \textbf{\bibinfo{volume}{102}}, \bibinfo{pages}{3477} (\bibinfo{year}{1995}).

\bibitem[{\citenamefont{L\"owdin and Shull}(1956)}]{LSH1956}
\bibinfo{author}{\bibfnamefont{P.-O.} \bibnamefont{L\"owdin}} \bibnamefont{and}
  \bibinfo{author}{\bibfnamefont{H.}~\bibnamefont{Shull}},
  \bibinfo{journal}{Phys. Rev.} \textbf{\bibinfo{volume}{101}},
  \bibinfo{pages}{1730} (\bibinfo{year}{1956}).

\bibitem[{\citenamefont{Theophilou et~al.}(2007)\citenamefont{Theophilou,
  Thanos, and Theophilou}}]{spin_contamination}
\bibinfo{author}{\bibfnamefont{I.}~\bibnamefont{Theophilou}},
  \bibinfo{author}{\bibfnamefont{S.}~\bibnamefont{Thanos}}, \bibnamefont{and}
  \bibinfo{author}{\bibfnamefont{A.~K.} \bibnamefont{Theophilou}},
  \bibinfo{journal}{The Journal of Chemical Physics}
  \textbf{\bibinfo{volume}{127}}, \bibinfo{eid}{234103} (\bibinfo{year}{2007}).

\bibitem[{\citenamefont{Atkin and Paula}(2014)}]{Atkins}
\bibinfo{author}{\bibfnamefont{P.}~\bibnamefont{Atkin}} \bibnamefont{and}
  \bibinfo{author}{\bibfnamefont{J.}~\bibnamefont{Paula}},
  \emph{\bibinfo{title}{Physical Chemistry}} (\bibinfo{publisher}{Oxford
  University Press}, \bibinfo{year}{2014}).

\bibitem[{\citenamefont{Schmidt et~al.}(1993)\citenamefont{Schmidt, Baldridge,
  Boatz, Elbert, Gordon, Jensen, Koseki, Matsunaga, Nguyen, Su
  et~al.}}]{GAMESS}
\bibinfo{author}{\bibfnamefont{M.~W.} \bibnamefont{Schmidt}},
  \bibinfo{author}{\bibfnamefont{K.~K.} \bibnamefont{Baldridge}},
  \bibinfo{author}{\bibfnamefont{J.~A.} \bibnamefont{Boatz}},
  \bibinfo{author}{\bibfnamefont{S.~T.} \bibnamefont{Elbert}},
  \bibinfo{author}{\bibfnamefont{M.~S.} \bibnamefont{Gordon}},
  \bibinfo{author}{\bibfnamefont{J.~J.} \bibnamefont{Jensen}},
  \bibinfo{author}{\bibfnamefont{S.}~\bibnamefont{Koseki}},
  \bibinfo{author}{\bibfnamefont{N.}~\bibnamefont{Matsunaga}},
  \bibinfo{author}{\bibfnamefont{K.~A.} \bibnamefont{Nguyen}},
  \bibinfo{author}{\bibfnamefont{S.}~\bibnamefont{Su}}, \bibnamefont{et~al.},
  \bibinfo{journal}{J.~Comput.~Chem.} \textbf{\bibinfo{volume}{14}},
  \bibinfo{pages}{1347} (\bibinfo{year}{1993}).

\bibitem[{cod()}]{code}
\bibinfo{note}{HIPPO computer program, info: lathiot@eie.gr}.

\end{thebibliography}

\end{document}